\documentclass[amsmath, pra, twocolumn, showpacs]{revtex4-1}
\usepackage{graphicx}
\usepackage{dsfont}

\begin{document}   

\title{Electromagnetically induced transparency for guided light in an atomic array outside an optical nanofiber}
 
\author{Fam Le Kien}

\affiliation{Wolfgang Pauli Institute, Oskar Morgensternplatz 1, 1090 Vienna, Austria}

\author{A. Rauschenbeutel} 

\affiliation{Vienna Center for Quantum Science and Technology, Institute of Atomic and Subatomic Physics, Vienna University of Technology, Stadionallee 2, 1020 Vienna, Austria}

\date{\today}

\begin{abstract}
We study the propagation of guided light along an array of three-level atoms in the vicinity of an optical nanofiber under the condition of electromagnetically induced transparency. We examine two schemes of atomic levels and field polarizations where the guided probe field is quasilinearly polarized along the major or minor principal axis, which is parallel or perpendicular, respectively, to the radial direction of the atomic position. Our numerical calculations indicate that 200 cesium atoms in a linear array with a length of 100 $\mu$m at a distance of 200 nm from the surface of a nanofiber with a radius of 250 nm can slow down the speed of guided probe light by a factor of about $3.5\times 10^6$ (the corresponding group delay is about 1.17 $\mu$s). In the neighborhood of the Bragg resonance, a significant fraction of the guided probe light can be reflected back with a negative group delay. The reflectivity and the group delay of the reflected field do not depend on the propagation direction of the probe field. However, when the input guided light is quasilinearly polarized along the major principal axis, the transmittivity and the group delay of the transmitted field substantially depend on the propagation direction of the probe field. Under the Bragg resonance condition, an array of atoms prepared in an appropriate internal state can transmit guided light polarized along the major principal in one specific direction even in the limit of infinitely large atom numbers. The directionality of transmission of guided light through the array of atoms is a consequence of the existence of a longitudinal component of the guided light field as well as the ellipticity of both the field polarization and the atomic dipole vector.
\end{abstract}

\maketitle

\section{Introduction}
\label{sec:introduction}

Optical properties of materials can be dramatically modified by quantum interference between the excitation pathways \cite{Harris review,Scully review}.
Intensive attention has been devoted to the manipulation and control of
the propagation of light through coherently driven optical media, especially in the connection with 
the possibility of enormous slowing down, storage, and retrieval of optical pulses \cite{Harris review,Scully review,slow light review,most recent review,Scully,Agarwal book,vapor,ultracold,polariton}. 
The interest to this topic is related to the applications for optical delay lines, optical data storage, optical memories, 
quantum computing, and sensitive measurements. Through the technique of electromagnetically induced transparency (EIT) \cite{Harris review,Scully review,slow light review,most recent review,Scully,Agarwal book}, ultralow group velocities of light have been obtained in hot \cite{vapor} and cold \cite{ultracold} atomic gases. This technique allows one to render the material highly transparent and still retain the strong dispersion required for the generation of slow light. In addition, the transmitted pulse can experience strong nonlinear effects due to the constructive interference in 
the third-order susceptibility $\chi^{(3)}$. This leads to new techniques in nonlinear optics at the few-photon level, which may find important applications to quantum information processing.

The EIT technique has been extended to media embedded in a variety of waveguide systems, such as rectangular waveguides \cite{Shen}, hollow-core photonic-crystal fibers \cite{Andre,Ghosh,Bajcsy09}, coupled resonator optical waveguides \cite{Neff}, quantum well waveguides \cite{Li}, waveguide-cavity systems \cite{Bermel}, and optical nanofibers \cite{Patnaik,propag,fibercavity,Spillane,Laurat15,Sayrin15a}. EIT-based photon switches in waveguides \cite{Bajcsy09,Ginzburg} have been examined. The generation and waveguiding of solitons in an EIT medium have been studied \cite{Hong}. Waveguiding of ultraslow light in an atomic Bose-Einstein condensate \cite{Tarhan} has been investigated. 

Nanofibers are optical fibers that are tapered to a diameter comparable to or smaller than the wavelength of light \cite{Mazur's Nature,Birks,taper}. Slowing down of guided light in an optical nanofiber embedded in an EIT medium has been investigated \cite{Patnaik,propag,fibercavity,Spillane,Laurat15,Sayrin15a}. 
The first observation of EIT at very low power levels of guided pump and probe light was reported in Ref.~\cite{Spillane}.
Very recently, coherent storage of guided light has been experimentally demonstrated \cite{Laurat15,Sayrin15a}.
In Ref. \cite{Patnaik}, the propagation of light is described in terms of the averages of the local refractive index, the local absorption coefficient, and the local group delay in the fiber cross-section plane. The studies in Refs. \cite{propag,fibercavity} are based on a more rigorous formalism that takes into account the inhomogeneous density distribution of the atomic gas and the inhomogeneous mode-profile function of the guided field in the fiber transverse plane. The atomic medium considered in \cite{Patnaik,propag,fibercavity,Spillane,Laurat15} is continuous. Meanwhile, recent experiments with atom-waveguide interfaces \cite{Dawkins11,Reitz13,Reitz14,Mitsch14a,Mitsch14b,Sayrin15b} used linear arrays of atoms prepared in a nanofiber-based optical dipole trap \cite{Vetsch10,Goban12,Polzik14}. 
It has recently been demonstrated experimentally that spin-orbit coupling of guided light can lead to directional spontaneous emission \cite{Mitsch14b,Petersen14} and optical diodes \cite{Sayrin15b}.
When the array period is near to the Bragg resonance, the discreteness and periodicity of the array may lead to significant effects, such as nearly perfect atomic mirrors, photonic band gaps,
long-range interaction, and self-ordering \cite{Deutsch95,Birkl95,Henkel03,Artoni05,Petrosyan07,Schilke11,Schilke12,Chang07,Chang11,Chang12,Ritsch14a,Ritsch14b,AtomArray}. 
In the prior work on atoms trapped in a one-dimensional optical lattice under the EIT condition \cite{Petrosyan07,Schilke12}, scalar light fields in free space were considered. Scattering of a scalar light field from an array of three-level atoms with two degenerate lower levels in a waveguide has also been studied \cite{Tsoi09}. 

In a nanofiber, the guided field penetrates an appreciable distance into the surrounding medium and appears as an evanescent wave carrying a significant fraction of the propagation power and having a complex polarization pattern \cite{fiber books,fibermode}. Since the nanofiber is thin, the guided modes of the nanofiber are the fundamental HE$_{11}$ modes  \cite{fiber books,fibermode}. These modes are hybrid modes. The field in such a mode has longitudinal electric and magnetic components. The local polarization of the mode varies in the fiber cross-section plane \cite{fibermode} and depends on the propagation direction \cite{Mitsch14b,Petersen14}. Therefore, the use of the scalar field formalism to treat the interaction of a nanofiber-guided field with an atom is not always appropriate. When the local polarization of the guided field is elliptical and the dipole matrix-element vector of the atom is a complex vector, direction-dependent effects in the atom-field interaction may occur \cite{Mitsch14b,Petersen14,Sayrin15b,AtomArray,Fam14}. 

In view of the recent results and insights, it is necessary to develop a systematic theory for the propagation of guided light under the EIT condition in an atomic array taking into account the vector nature of the guided field and the discreteness and periodicity of the array. 

In this paper, we study EIT in a one-dimensional periodic array of three-level atoms trapped along an optical nanofiber. 
We examine two schemes of atomic levels and field polarizations where the guided probe field is quasilinearly polarized along the major principal axis $x$ or the minor principal axis $y$, which lie in the fiber cross-section plane and are parallel or perpendicular, respectively, to the radial direction of the atomic position. We take into account the vector nature of the guided field and the discreteness and periodicity of the atomic array. We study the transmittivity and reflectivity of guided light and the time evolution of the transmitted and reflected fields.

The paper is organized as follows. In Sec.~\ref{sec:model} we describe two schemes of atomic levels and field polarizations for nanofiber-based EIT, and present the coupled-mode propagation equations. In Sec.~\ref{sec:continuous_medium} we study EIT in the homogeneous-medium approximation and the phase-matching approximation. 
In Sec.~\ref{sec:array} we investigate EIT in a discrete array of atoms with the help of the transfer matrix formalism. Our conclusions are given in Sec.~\ref{sec:summary}.

\section{Nanofiber-based EIT schemes and coupled-mode propagation equations}
\label{sec:model}

\begin{figure}[tbh]
\begin{center}
  \includegraphics{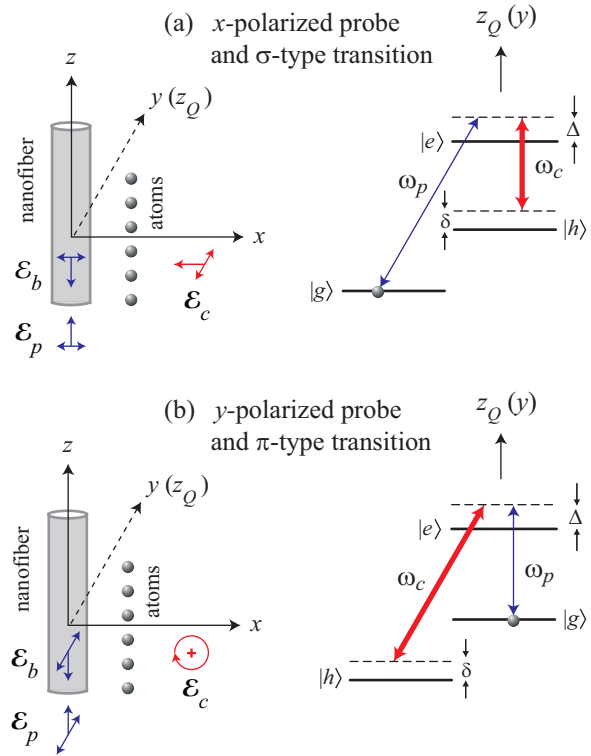}
\end{center}
\caption{(Color online)
Left column: An array of atoms outside an optical nanofiber with a plane-wave control field $\boldsymbol{\mathcal{E}}_c$, a guided probe field $\boldsymbol{\mathcal{E}}_p$, and a guided reflected field $\boldsymbol{\mathcal{E}}_b$. The atomic array is aligned parallel to the fiber axis $z$ and lies in the $zx$ plane of the Cartesian coordinate frame $\{x,y,z\}$. The probe field $\boldsymbol{\mathcal{E}}_p$ is quasilinearly polarized along the major principal axis $x$ in scheme (a) and the minor principal axis $y$ in scheme (b). 
The control field $\boldsymbol{\mathcal{E}}_c$ propagates along the axis $x$ and is linearly polarized along the axis $y$ in scheme (a), and propagates along the axis $y$ and is circularly polarized in scheme (b). Right column: The diagram of atomic energy levels and electric dipole transitions for the analysis. Atomic cesium is used. The quantization axis $z_Q$ is the axis $y$. In both schemes (a) and (b), the upper level is $|e\rangle=|6P_{3/2},F'=4, M'=\pm 4\rangle$. In scheme (a), the lower levels are $|g\rangle=|6S_{1/2},F=3, M=\pm 3\rangle$ and $|h\rangle=|6S_{1/2},F=4, M=\pm 4\rangle$. In scheme (b), the lower levels are $|g\rangle=|6S_{1/2},F=4, M=\pm 4\rangle$ and $|h\rangle=|6S_{1/2},F=3, M=\pm 3\rangle$.     
}
\label{fig1}
\end{figure} 

We consider a linear periodic array of $\Lambda$-type three-level atoms trapped outside an optical nanofiber (see Fig. \ref{fig1}).
The nanofiber has a cylindrical silica core, with radius $a$ and refractive index $n_1$, surrounded by vacuum, with refractive index $n_2=1$. 
We use the Cartesian coordinate system $\{x,y,z\}$ and the associated cylindrical coordinate system $\{r,\varphi,z\}$, with $z$ being the fiber axis. We assume that the array of atoms is parallel to the fiber axis $z$ and lies in the $zx$ plane. The positions of the atoms in the array are characterized by the Cartesian coordinates $x_j=x_0$, $y_j=0$, and $z_j=(j-1)\Lambda$. Here, the index $j=1,2,\dots,N$ labels the atoms, with $N$ being the number of atoms in the array, and the parameter $\Lambda$ is the period of the array. 
The axes $x$ and $y$, which lie in the fiber cross-section plane and are parallel and perpendicular, respectively, to the radial direction of the atomic position, are called
the major and minor principal axes, respectively. Although our theory is general and applicable, in principle, to arbitrary multilevel atoms, we assume cesium atoms throughout this paper. For simplicity, we neglect the effect of the surface-induced potential on the atomic energy levels. This approximation is reasonable when the atoms are not close to the fiber surface.
We also neglect the effect of the far-detuned trapping light fields.

\subsection{Quasilinearly polarized nanofiber-guided modes}

We represent the electric component of a nanofiber-guided light field as
$\mathbf{E}=(\boldsymbol{\mathcal{E}}e^{-i\omega t}+\mathrm{c.c.})/2
=(\mathcal{E}\mathbf{u}e^{-i\omega t}+\mathrm{c.c.})/2$,
where $\omega$ is the angular frequency and $\boldsymbol{\mathcal{E}}=\mathcal{E}\mathbf{u}$ is the slowly varying envelope 
of the positive-frequency part,
with $\mathcal{E}$ and $\mathbf{u}$ being the field amplitude and the polarization vector, respectively.
In general, the amplitude $\mathcal{E}$ is a complex scalar and the polarization vector $\mathbf{u}$ is a complex unit vector. 
The guided light field can be decomposed into a superposition of quasilinearly polarized modes \cite{fiber books}.
These guided modes can be labeled by the index $(\omega,f,\xi)$, where 
$f=+1$ or $-1$ (or simply $+$ or $-$) stands for the positive ($+\hat{\mathbf{z}}$) or negative ($-\hat{\mathbf{z}}$) propagation direction, respectively, and 
$\xi=x$ or $y$ stands for the major polarization axis. 
In the cylindrical coordinates, the transverse-plane profile functions of the positive-frequency parts of  
the electric components of the modes $(\omega,f,\xi)$ are given by \cite{AtomArray,Fam14,fiber books,fibermode}
\begin{equation}\label{g9}
\begin{split}
\mathbf{e}^{(\omega fx)}=&\sqrt2\,(\hat{\mathbf{r}}e_r\cos\varphi+i\hat{\boldsymbol{\varphi}}e_\varphi\sin\varphi+f\hat{\mathbf{z}}e_z\cos\varphi),\\
\mathbf{e}^{(\omega fy)}=&\sqrt2\,(\hat{\mathbf{r}}e_r\sin\varphi-i\hat{\boldsymbol{\varphi}}e_\varphi\cos\varphi+f\hat{\mathbf{z}}e_z\sin\varphi).
\end{split}
\end{equation}
Here, the notations $\hat{\mathbf{r}}=\hat{\mathbf{x}}\cos\varphi + \hat{\mathbf{y}}\sin\varphi$, 
$\hat{\boldsymbol{\varphi}}=-\hat{\mathbf{x}}\sin\varphi + \hat{\mathbf{y}}\cos\varphi$, and $\hat{\mathbf{z}}$
stand for the unit basis vectors of the cylindrical coordinate system, where $\hat{\mathbf{x}}$ and $\hat{\mathbf{y}}$ are the unit basis vectors of the Cartesian coordinate system for the fiber cross-section plane $xy$. The notations $e_r=e_r(r)$, $e_{\varphi}=e_{\varphi}(r)$, and $e_z=e_z(r)$ stand for
the cylindrical components of the profile function $\mathbf{e}^{(\omega,+,+)}(r,\varphi)$ of the forward counterclockwise polarized guided mode and are given in Refs.~\cite{fiber books,fibermode,cesium decay,AtomArray,Fam14}. Equations \eqref{g9} show that the $x$- and $y$-polarized guided modes have, in general, not only transverse but also longitudinal components. The local polarizations of these modes vary in the fiber cross-section plane, and are generally not strictly linear \cite{fiber books,fibermode}. It is interesting to note from Eqs.~\eqref{g9} that the signs of the longitudinal components $fe_z\cos\varphi$ and $fe_z\sin\varphi$ for the $x$- and $y$-polarized modes, respectively, depend on the propagation direction $f$. Thus, the difference between the mode profile functions for the forward ($f=+$) and backward ($f=-$) guided fields is expressed by the change in sign of the longitudinal components. This change may affect the magnitude of the coupling between the atom and the field and, consequently, may lead to directional spontaneous emission and directional scattering \cite{Fam14,AtomArray,Mitsch14b,Petersen14,Sayrin15b}.

Since the radial direction of the atomic position in our study is parallel to the axis $x$, the polar angle for the atomic position is $\varphi=0$. Therefore, when we evaluate the mode functions \eqref{g9} at the positions of the atoms, we find
\begin{subequations}\label{g10}
\begin{eqnarray}
\mathbf{e}^{(\omega fx)}|_{x \text{ axis}}&=&\sqrt2\,(i|e_r|\hat{\mathbf{x}}+f|e_z|\hat{\mathbf{z}}),\label{g10a}\\
\mathbf{e}^{(\omega fy)}|_{x \text{ axis}}&=&i\sqrt2\,|e_\varphi|\hat{\mathbf{y}}.\label{g10b}
\end{eqnarray}
\end{subequations}
In deriving the above equations we have used the properties $e_r=i|e_r|$, 
$e_\varphi=-|e_\varphi|$,  and $e_z=|e_z|$ \cite{AtomArray,Fam14,fiber books,fibermode}.

Equation \eqref{g10a} indicates that, on the $x$ axis,
the $x$-polarized guided mode has two components, the transverse component $i\sqrt2\,|e_r|$ (aligned along the $x$ axis) 
and the longitudinal component $f\sqrt2\,|e_z|$ (aligned along the $z$ axis).
The difference in phase between these components is $f\pi/2$. It depends of the mode propagation direction $f$.
The polarization vector of the $x$-polarized guided mode on the $x$ axis is  
\begin{equation}\label{g11} 
\mathbf{u}=\frac{i|e_r|\hat{\mathbf{x}}+f|e_z|\hat{\mathbf{z}}}
{\sqrt{|e_r|^2+|e_z|^2}}. 
\end{equation}
It is clear that the local polarization of the $x$-polarized guided mode on the $x$ axis is elliptical in the $zx$ plane.  
The ellipticity vector is given by $i[\mathbf{u}\times \mathbf{u}^*]=\sigma_{\mathrm{ell}}\hat{\mathbf{y}}$, where 
\begin{equation}\label{g12} 
\sigma_{\mathrm{ell}}=f\frac{2|e_r||e_z|}{|e_r|^2+|e_z|^2}
\end{equation}
is the ellipticity parameter.
The circulation direction of the above elliptical polarization depends on the mode propagation direction $f$. 
The polarization vector \eqref{g11} is a linear superposition of 
the circular polarization basis vectors $\boldsymbol{\sigma}^+=(\hat{\mathbf{z}}+i\hat{\mathbf{x}})/\sqrt2$ 
and $\boldsymbol{\sigma}^-=-(\hat{\mathbf{z}}-i\hat{\mathbf{x}})/\sqrt2$ with the coefficients
\begin{equation}\label{g13}
\begin{split}  
u_{\sigma^+}&=\frac{1}{\sqrt{2}}\frac{|e_r|+f|e_z|}{\sqrt{|e_r|^2+|e_z|^2}},\\ 
u_{\sigma^-}&=\frac{1}{\sqrt{2}}\frac{|e_r|-f|e_z|}{\sqrt{|e_r|^2+|e_z|^2}}.
\end{split}
\end{equation}
The corresponding weight factors $|u_{\sigma^\pm}|^2$ are given in terms of the polarization ellipticity parameter $\sigma_{\mathrm{ell}}$ as $|u_{\sigma^\pm}|^2=(1\pm\sigma_{\mathrm{ell}})/2$.
When $|e_z| \simeq |e_r|$, we have $\sigma_{\mathrm{ell}}\simeq\pm1$, $|u_{\sigma^\pm}|^2\simeq 1$, and $|u_{\sigma^\mp}|^2\simeq 0$ for $f=\pm$. In this case, the local polarization of the $x$-polarized guided mode at the position of the atom on the $x$ axis is almost circular. As an example, we consider the case where the nanofiber radius is $a=250$ nm and the atom-to-surface distance is $r-a=200$ nm. These parameters correspond to the Vienna atom trap experiment \cite{Vetsch10}. 
We find the ratio $|e_z|/|e_r|\simeq0.55$, which leads to $\sigma_{\mathrm{ell}}\simeq\pm0.84$, 
$|u_{\sigma^\pm}|^2\simeq 0.92$, and $|u_{\sigma^\mp}|^2\simeq 0.08$ for $f=\pm$ \cite{Mitsch14b,Petersen14}. 

Equation \eqref{g10b} indicates that, on the $x$ axis, the $y$-polarized guided mode has a single component $i\sqrt2\,|e_\varphi|$, which is aligned along the $y$ axis. Thus, the local polarization of the $y$-polarized guided mode at the position of the atomic array is exactly linear along the $y$ axis. This local polarization does not depends on the mode propagation direction $f$.  

\subsection{Atomic levels and EIT schemes}

We assume that the atoms have a single upper level $|e\rangle$ of energy $\hbar\omega_e$ and two lower levels $|g\rangle$ and $|h\rangle$ of energies $\hbar\omega_g$ and $\hbar\omega_h$, respectively.  
The atoms are initially prepared in the lower level $|g\rangle$. The levels $|e\rangle$ and $|g\rangle$ are coupled by a weak probe field $\boldsymbol{\mathcal{E}}_p$ of frequency $\omega_p$. The levels $|e\rangle$ and $|h\rangle$ are coupled by a strong control field $\boldsymbol{\mathcal{E}}_{c}$ of frequency $\omega_c$. The transition between the lower levels $|g\rangle$ and $|h\rangle$ is electric-dipole forbidden. We assume that the probe field $\boldsymbol{\mathcal{E}}_p$ is guided by the nanofiber and propagates along the fiber axis $z$ in the direction $f_p=\pm$, where $+$ or $-$
corresponds to the positive direction $+\hat{\mathbf{z}}$ or the negative direction $-\hat{\mathbf{z}}$, respectively. The detuning of the probe field with respect to the atomic transition $|e\rangle\leftrightarrow |g\rangle$ is denoted by
\begin{equation}\label{e1}
\Delta =\omega_p-\omega_0, 
\end{equation}
where $\omega_0=\omega_{eg}\equiv\omega_{e}-\omega_{g}$.
The control field $\boldsymbol{\mathcal{E}}_{c}$ is an external plane-wave field propagating perpendicularly to the fiber axis $z$. 
The two-photon (Raman) transition between the lower levels $|g\rangle$ and $|h\rangle$ may be off resonance, and the corresponding two-photon detuning is denoted by 
\begin{equation}\label{e2}
\delta =\omega_p-\omega_{c}-\omega_{h}+\omega_{g}. 
\end{equation}
In our general analytical calculations, the single-photon detuning $\Delta$ and the two-photon detuning $\delta$ can be different from each other.
However, in our numerical calculations, we will study only the case where $\Delta=\delta$, that is, the case where the control field is on exact resonance with the atomic transition
$|e\rangle\leftrightarrow |h\rangle$.

To be specific, we use the transitions between the Zeeman sublevels of the $D_2$ line of atomic cesium in our calculations. 
In order to specify the internal atomic states, we use the minor principal axis $y$ as the quantization axis $z_Q$. 
The purpose of this choice is that it allows us to identify appropriate atomic states $|e\rangle$ and $|g\rangle$  which are coupled to each other
by just one type of polarization of guided probe light. Indeed, as shown in the previous subsection, at an arbitrary position on the axis $x$, the local polarization of $x$-polarized guided light is elliptical in the $zx$ plane and the local polarization of $y$-polarized guided light is exactly linear along the $y$ axis. Therefore, an atomic transition between two Zeeman sublevels specified with respect to the quantization axis $y$ can interact with either $x$- or $y$-polarized guided light but not with $y$- or $x$-polarized guided light, respectively.

We consider two schemes that are illustrated in parts (a) and (b) of Fig.~\ref{fig1}. In both schemes, 
we use the excited-state sublevel $|6P_{3/2},F'=4, M'=\pm 4\rangle$ as the upper level $|e\rangle$. Furthermore, we use the ground-state sublevels $|6S_{1/2},F=3, M=\pm 3\rangle$ and $|6S_{1/2},F=4, M=\pm 4\rangle$ as the levels $|g\rangle$ and $|h\rangle$, respectively, in scheme (a) of Fig.~\ref{fig1}, and as the levels
$|h\rangle$ and $|g\rangle$, respectively, in scheme (b) of Fig.~\ref{fig1}. The effects of other Zeeman sublevels are removed by applying an external magnetic field.
We emphasize that the atomic states in the above schemes are specified  by using the minor principal axis $y$ as the quantization axis.
In addition, the quantum numbers of the lower states $|g\rangle$ and $|h\rangle$ are interchangeable between the two different schemes (see Fig.~\ref{fig1}). 

In scheme (a) of Fig.~\ref{fig1}, the guided probe field $\boldsymbol{\mathcal{E}}_p$ is quasilinearly polarized along the $x$ direction. Meanwhile, the control field $\boldsymbol{\mathcal{E}}_{c}$ is a plane wave propagating along the axis $x$ and linearly polarized along the axis $y=z_Q$ in accordance with the $\pi$ type of the atomic transition $|e\rangle\leftrightarrow|h\rangle$. In scheme (b) of Fig.~\ref{fig1}, the guided probe field $\boldsymbol{\mathcal{E}}_p$ is quasilinearly polarized along the $y$ direction. Meanwhile, the control field $\boldsymbol{\mathcal{E}}_{c}$ is a plane wave propagating along the axis $y$ and counterclockwise or clockwise circularly polarized in accordance with the $\sigma_{+}$ or $\sigma_{-}$ type of the atomic transition $|e\rangle\leftrightarrow|h\rangle$. In what follows, schemes (a) and (b) of Fig.~\ref{fig1} are called the $x$- and $y$-polarization schemes, respectively. In both schemes, we neglect the reflection of the control field $\boldsymbol{\mathcal{E}}_{c}$ from the fiber surface. Since the control field propagation direction and the atomic array axis are perpendicular and parallel, respectively, to the fiber axis $z$, the effect of the reflection
of the control field $\boldsymbol{\mathcal{E}}_{c}$ can be easily accounted for by modifying the magnitude of $\boldsymbol{\mathcal{E}}_{c}$ at the position of the atomic array.

We introduce the notation $M_\alpha$ for the magnetic quantum number of the atomic level $\alpha$, where $\alpha=e,g,h$. In the analytical calculations, we use $M_g=\pm3$ and  $M_h=\pm4$ for the $x$-polarization scheme, $M_g=\pm4$ and  $M_h=\pm3$ for the $y$-polarization scheme, and $M_e=\pm4$ for both schemes.  

We note that, in both schemes (a) and (b) of Fig.~\ref{fig1}, the probe transition $|e\rangle\leftrightarrow|g\rangle$ is not coupled
to the guided modes with the polarization $\bar{\xi}_p$ that is orthogonal to the polarization $\xi_p$ of the incident guided probe field.
Here, we have introduced the notation $\bar{\xi}=y$ for $\xi=x$ and $\bar{\xi}=x$ for $\xi=y$. 
For the control field $\boldsymbol{\mathcal{E}}_{c}$, we can use a guided field instead of an external plane-wave field. Indeed, the control transition $|e\rangle\leftrightarrow|h\rangle$ can be coupled by an additional guided field that is quasilinearly polarized along the $y$ direction in scheme (a) of Fig.~\ref{fig1} or along the $x$ direction in scheme (b) of the figure. 

Due to the interaction between the atoms and the guided probe field $\boldsymbol{\mathcal{E}}_p$, a guided reflected field $\boldsymbol{\mathcal{E}}_b$ with the frequency $\omega_b=\omega_p$, the propagation direction $f_b=-f_p$, and the polarization $\xi_b=\xi_p$ may be generated. In order to describe the reflection, we need to include both propagation directions $f=+$ and $f=-$ into the analysis. 
We introduce the notation $\boldsymbol{\mathcal{E}}_f$ for the positive frequency component of
the electric part of the field in the guided mode with the frequency $\omega_p$, the polarization $\xi_p$, and the propagation direction $f=\pm$.
The electric field vector $\boldsymbol{\mathcal{E}}_f=\boldsymbol{\mathcal{E}}_f(r,\varphi,z,t)$ is related to the photon flux amplitude
$\mathcal{A}_f=\mathcal{A}_f(z,t)$ via the formula
\begin{equation}\label{x15}
\boldsymbol{\mathcal{E}}_f=i\sqrt{\frac{2\hbar\omega_p}{\epsilon_0v_g}}
\mathcal{A}_f\mathbf{e}^{(\omega_pf\xi_p)}.
\end{equation}
Here, $v_g=(d\beta/d\omega)^{-1}|_{\omega=\omega_p}$ is the group velocity of the guided field, where $\beta$ is the longitudinal propagation constant. The notation $\mathbf{e}^{(\omega f \xi)}$ stands for the normalized profile function for the guided mode with the frequency $\omega$, the propagation direction $f$, and the polarization $\xi$ \cite{fiber books,cesium decay,AtomArray,Fam14}.
According to \cite{propag}, the amplitudes $\mathcal{A}_f=\mathcal{A}_{\pm}$ of the photon fluxes of the guided fields $\boldsymbol{\mathcal{E}}_f=\boldsymbol{\mathcal{E}}_{\pm}$
are governed, in the framework of the slowly varying envelope approximation, by the propagation equations
\begin{eqnarray}\label{e9}
\left(\frac{\partial }{\partial z}-i\beta_p\right)\mathcal{A}_{+} &=& n_A G^*_{+} \rho_{eg},\nonumber\\ 
\left(\frac{\partial }{\partial z}+i\beta_p\right)\mathcal{A}_{-} &=& -n_A G^*_{-} \rho_{eg}. 
\end{eqnarray}
Here, $\beta_p=\beta(\omega_p)$ is the longitudinal propagation constant for the forward and backward guided fields, 
\begin{equation}\label{e9a}
n_A=\sum_j \delta(z-z_j)
\end{equation} 
is the one-dimensional atom-number density, 
$G_{+}$ and $G_{-}$ are the coupling coefficients for the guided fields $\boldsymbol{\mathcal{E}}_{+}$ and $\boldsymbol{\mathcal{E}}_{-}$, respectively,
and $\rho_{\alpha\alpha'}\equiv \langle \alpha|\rho|\alpha'\rangle$ with $\alpha,\alpha'=e,g,h$ are the elements of the density matrix $\rho$ the atom in the
interaction picture.
The coupling coefficient $G_f=G_{\pm}$ is defined as 
\begin{equation}\label{e10}
G_f=\sqrt{\frac{\omega_p}{2\epsilon_0\hbar v_g}}\;
\mathbf{d}_{eg}\cdot\mathbf{e}^{(\omega_pf\xi_p)}.
\end{equation}
Here, $\mathbf{d}_{eg}$ is the dipole matrix element for the atomic transition $|e\rangle\leftrightarrow|g\rangle$.
We emphasize that the photon flux amplitudes $\mathcal{A}_{\pm}(z,t)$ are independent of $x$ and $y$.  

It is clear from Eq.~\eqref{e10} that the atom-field coupling coefficient $G_f$ depends on the local polarization of the guided probe field at the position of the atom.
This coefficient also depends on the orientation and magnitude of the dipole matrix element $\mathbf{d}_{eg}$.
We emphasize again that, in our study, the internal atomic states and the atomic transitions are specified  by using the minor principal axis $y$ as the quantization axis.
Moreover, in order to obtain nonzero atom-field coupling coefficients, different transitions $|e\rangle\leftrightarrow|g\rangle$ of atomic cesium, with different dipole matrix elements $\mathbf{d}_{eg}$, are used in the different polarization schemes (see Fig.~\ref{fig1}). 

In the case of the $x$-polarization scheme, the coupling coefficient $G_f$ is given as
\begin{equation}\label{e12}
G_f=-\sqrt{\frac{\omega_p}{2\epsilon_0\hbar v_g}}\;
d_{eg}(|e_r|+f(M_e-M_g) |e_z|).
\end{equation}
Here, $e_r$ and $e_z$ are respectively the radial and axial components of the mode profile function $\mathbf{e}\equiv\mathbf{e}^{(\omega_p,+,+)}$ of the forward counterclockwise quasicircularly polarized guided modes \cite{fiber books,cesium decay,AtomArray,Fam14}.

In the case of the $y$-polarization scheme, the coupling coefficient $G_f$ is given as
\begin{eqnarray}\label{e11}
G_f=i\sqrt{\frac{\omega_p}{\epsilon_0\hbar v_g}}\; d_{eg}|e_\varphi|.
\end{eqnarray}
Here, $e_\varphi$ is the azimuthal component of the mode profile function $\mathbf{e}\equiv\mathbf{e}^{(\omega_p,+,+)}$ of the forward counterclockwise quasicircularly polarized guided modes \cite{fiber books,cesium decay,AtomArray,Fam14}.

Note that $|G_f|^2=\gamma_{eg}^{(f\xi_p)}$, where $\gamma_{eg}^{(f\xi_p)}$ is the rate of spontaneous emission of the atomic transition $|e\rangle\leftrightarrow|g\rangle$ into the guided mode with the propagation direction $f$ and the polarization $\xi_p$. Since the probe transition $|e\rangle\leftrightarrow|g\rangle$ is not coupled to the guided modes with the polarization $\bar{\xi}_p$ in the cases of schemes (a) and (b) of Fig.~\ref{fig1}, we have the relation $\gamma_{eg}^{(f)}=\gamma_{eg}^{(f\xi_p)}$, where $\gamma_{eg}^{(f)}$ is the rate of spontaneous emission of the atomic transition $|e\rangle\leftrightarrow|g\rangle$ into the guided modes with the propagation direction $f$. Hence, we obtain $|G_f|^2=\gamma_{eg}^{(f)}$. 

The Rabi frequency caused by the guided field $\boldsymbol{\mathcal{E}}_f$ is given by 
\begin{equation}\label{e15a}
\Omega_f=\mathbf{d}_{eg}\cdot\boldsymbol{\mathcal{E}}_f/\hbar=2iG_f\mathcal{A}_f.
\end{equation}
The Rabi frequency caused by the control field $\boldsymbol{\mathcal{E}}_c$ is $\Omega_c=\mathbf{d}_{eh}\cdot\boldsymbol{\mathcal{E}}_c/\hbar$, 
where $\mathbf{d}_{eh}$ is the dipole matrix element for the atomic transition $|e\rangle\leftrightarrow|h\rangle$.
We assume that $|\Omega_{\pm}|\ll |\Omega_c|$. In the adiabatic approximation,
the expression for $\rho_{eg}$ to first order in $\Omega_{\pm}$ is found to be
\begin{equation}\label{e15}
\rho_{eg}=\frac{\Omega_{+}+\Omega_{-}}{2}\mathcal{F},
\end{equation}
where \cite{Harris review,Scully review,slow light review,most recent review,Scully,Agarwal book}
\begin{equation}\label{e18}
\mathcal{F} = \frac{\delta+i\Gamma_{hg}}{|\Omega_c|^2/4-(\Delta+i\Gamma_{eg})(\delta+i\Gamma_{hg})}.
\end{equation}
Here, $\Gamma_{eg}$ is the decay rate of the atomic probe transition coherence $\rho_{eg}$, and $\Gamma_{hg}$ is the decay rate of the lower-level coherence $\rho_{hg}$. 

We insert Eq.~\eqref{e15} into Eqs.~\eqref{e9} and make use of expression \eqref{e15a}. Then, we obtain
\begin{eqnarray}\label{e16}
\left(\frac{\partial }{\partial z}-i\beta_p\right)\mathcal{A}_{+}&=&i(\tilde{K}_{++}\mathcal{A}_{+}+\tilde{K}_{+-}\mathcal{A}_{-}),\nonumber\\
\left(\frac{\partial }{\partial z}+i\beta_p\right)\mathcal{A}_{-}&=&-i(\tilde{K}_{-+}\mathcal{A}_{+}+\tilde{K}_{--}\mathcal{A}_{-}),
\end{eqnarray}
where
\begin{equation}\label{e17}
\tilde{K}_{ff'}=n_A G_{f}^*G_{f'}\mathcal{F}
\end{equation}
for $f,f'=+,-$. In general, the one-dimensional atom-number density $n_A$ and consequently the coupling coefficients $\tilde{K}_{ff'}$ are generalized functions of $z$ [see Eq.~\eqref{e9a}].

\section{Continuous-medium approximations}
\label{sec:continuous_medium}

In this section, we approximate the generalized-function representation \eqref{e9a} of the one-dimensional atom-number density $n_A$ by two different continuous-function representations and present the corresponding analytical and numerical results. In Sec.~\ref{subsec:homogeneous}, we replace $n_A$ by a constant and solve the corresponding coupled-mode propagation equations. 
In Sec.~\ref{subsec:phase-matching}, we expand $n_A$ into a Fourier series and neglect the terms that do not correspond to the phase-matching condition in the coupled-mode propagation equations. 

\subsection{Homogeneous-medium approximation}
\label{subsec:homogeneous}

We consider the case where the lattice constant $\Lambda$ is not close to any integer multiple of the in-fiber half-wavelength $\lambda_F/2=\pi/\beta_p$ of the probe field, that is, 
the atomic array is far off the Bragg resonance. 
In this case, the effect of the interference between the beams reflected from different atoms in the array is not significant and, therefore,
we can neglect the discreteness and periodicity of the atomic array. 
This approximation means that we can use the one-dimensional atom-number distribution 
\begin{equation}\label{e19b}
n_A=1/\Lambda,
\end{equation}
which is continuous and constant in the axial coordinate $z$.
With the use of this approximation, expression (\ref{e17}) for the coefficients $\tilde{K}_{ff'}$ reduces to 
\begin{equation}\label{e19a}
\tilde{K}_{ff'}=K_{ff'}, 
\end{equation}
where the coefficients
\begin{equation}\label{e19}
K_{ff'}=G_{f}^*G_{f'}\frac{\mathcal{F}}{\Lambda} 
\end{equation}
are independent of $z$. Note that $K_{+-}K_{-+}=K_{++}K_{--}$. 
In the case of the $x$-polarization scheme, we have $K_{++}\not=K_{--}$ and $K_{+-}=K_{-+}$. 
In the case of the $y$-polarization scheme, we have $K_{++}=K_{--}=K_{+-}=K_{-+}$. 
It is convenient to introduce the notations $K_{+}=K_{++}$ and $K_{-}=K_{--}$. 

We can easily solve Eqs.~\eqref{e16} with the constant coefficients given by Eq.~\eqref{e19a}.
We assume that $z=0$ and $z=L=(N-1)\Lambda$ are the left- and right-edge positions of the atomic medium, respectively.
In the case where the incident probe field is $\mathcal{A}_{+}(0)$, the boundary condition is $\mathcal{A}_{-}(L)=0$. In this case, 
the reflection and transmission coefficients are $R_A^{(+)}=\mathcal{A}_{-}(0)/\mathcal{A}_{+}(0)$ and $T_A^{(+)}=\mathcal{A}_{+}(L)/\mathcal{A}_{+}(0)$, respectively.
In the case where the incident probe field is $\mathcal{A}_{-}(L)$, the boundary condition is $\mathcal{A}_{+}(0)=0$. 
In this case, the reflection and transmission coefficients are $R_A^{(-)}=\mathcal{A}_{+}(L)/\mathcal{A}_{-}(L)$ and $T_A^{(-)}=\mathcal{A}_{-}(0)/\mathcal{A}_{-}(L)$, respectively.
With the help of the relation $K_{+-}=K_{-+}$, we can show that $R_A^{(+)}=R_A^{(-)}\equiv R_A$. 
The expressions for $R_A$ and $T_A^{(f)}$ are found to be
\begin{eqnarray}\label{e45}
R_A&=&\frac{iK_{-+}\sin(QL)}{Q\cos(QL)-i[\beta_p+(K_{+}+K_{-})/2]\sin(QL)},
\nonumber\\
T_A^{(f)}&=&\frac{Q\exp[if(K_{+}-K_{-})L/2]}{Q\cos(QL)-i[\beta_p+(K_{+}+K_{-})/2]\sin(QL)},\qquad
\end{eqnarray}
where
\begin{equation}\label{e21}
Q=\sqrt{\beta_p^2+\beta_p(K_{+}+K_{-})+\frac{(K_{+}-K_{-})^2}{4}}.
\end{equation}
We note that, in the case of the $y$-polarization scheme, we have $K_{+}=K_{-}$, which leads to 
$T_A^{(+)}=T_A^{(-)}\equiv T_A$. 

Since $|G_{f}^*G_{f'}|< 2\Gamma_{eg}$, $|\mathcal{F}|\leq 1/\Gamma_{eg}$, and $\Lambda\sim\pi/\beta_p$, we have $|K_{ff'}|\ll\beta_p$.
Hence, we find $Q\simeq \beta_p+(K_{+}+K_{-})/2$.
With this approximation, Eqs.~\eqref{e45} reduce to
\begin{eqnarray}\label{e28}
R_A&\simeq&\frac{K_{-+}}{2\beta_p+K_{+}+K_{-}}\{\exp [i(2\beta_p+K_{+}+K_{-})L]-1\},
\nonumber\\
T_A^{(f)}&\simeq&\exp [i(\beta_p+K_f) L].
\end{eqnarray}

Since $|K_{-+}|\ll\beta_p$, the reflectivity of the array is 
$|R_A|^2\simeq 0$. 
The transmission of the probe field is $|T_A^{(f)}|^2\simeq\exp(-2\kappa_f L)$, where
\begin{equation}\label{e32}
\kappa_f=\mathrm{Im}(K_f) 
\end{equation}
is the absorption coefficient for the probe field in the case where the reflection is negligible.
The corresponding phase shift coefficient for the probe field is
\begin{equation}\label{e33}
\theta_f=\mathrm{Re}(K_f).
\end{equation}
The optical depth per atom $\mathcal{D}_f\equiv 2\kappa_f\Lambda$ is
\begin{equation}\label{e32a}
\mathcal{D}_f=2|G_f|^2\mathrm{Im}\left(\frac{\delta+i\Gamma_{hg}}{|\Omega_c|^2/4-(\Delta+i\Gamma_{eg})(\delta+i\Gamma_{hg})}\right).
\end{equation}
The phase shift per atom $\Theta_f\equiv \theta_f\Lambda$ is 
\begin{equation}\label{e33a}
\Theta_f=|G_f|^2\mathrm{Re}\left(\frac{\delta+i\Gamma_{hg}}{|\Omega_c|^2/4-(\Delta+i\Gamma_{eg})(\delta+i\Gamma_{hg})}\right).
\end{equation}

We now describe time dependence of the guided probe field $\mathcal{A}_f$, where $f=f_p$. In the case where the period of the atomic array is far from the Bragg resonance, the reflection is, as shown analytically above and illustrated numerically in Figs.~\ref{fig4}(b) and \ref{fig5}(b) below, negligible. Then, in the frequency domain, the Fourier-transformed amplitude $\tilde{\mathcal{A}_f}(z,\omega)=(2\pi)^{-1/2}\int_{-\infty}^{\infty}\mathcal{A}_f(z,t)e^{i(\omega-\omega_p)t}dt$ of the probe field is governed by the propagation equation
\begin{eqnarray}\label{e34}
\frac{\partial \tilde{\mathcal{A}_f}(z,\omega)}{f\partial z}=i[\beta(\omega)+K_f(\omega)]\tilde{\mathcal{A}_f}(z,\omega).
\end{eqnarray}
Here, we have introduced the notation $K_f(\omega)=K_{ff}(\omega)$, where $K_{ff'}(\omega)=K_{ff'}|_{\omega_p=\omega}$ is given by Eq.~\eqref{e19}
with the substitution $\omega_p=\omega$.
We neglect the dispersion of the fiber-mode group velocity, that is, we take $\beta(\omega)=\beta_p+\beta'_p(\omega-\omega_p)$, where $\omega_p$ is the central frequency of the input guided probe field. 
We expand $K_f(\omega)$ up to the second order of $\omega-\omega_p$ as 
\begin{equation}\label{e35}
K_f(\omega)=K_f+K'_f (\omega-\omega_p)+\frac{1}{2}K''_f(\omega-\omega_p)^2,
\end{equation} 
where
\begin{widetext}
\begin{eqnarray}\label{e36}
K_f&=&\frac{ |G_f|^2}{\Lambda}\frac{\delta+i\Gamma_{hg}}{|\Omega_c|^2/4-(\Delta+i\Gamma_{eg})(\delta+i\Gamma_{hg})},
\nonumber \\
K'_f&=&\frac{ |G_f|^2}{\Lambda}\frac{|\Omega_c|^2/4+(\delta+i\Gamma_{hg})^2}
{\big[|\Omega_c|^2/4-(\Delta+i\Gamma_{eg})(\delta+i\Gamma_{hg})\big]^2},
\nonumber\\
K''_f&=&2\frac{ |G_f|^2}{\Lambda}
\frac{\big[\Delta+i\Gamma_{eg}+2(\delta+i\Gamma_{hg})\big]|\Omega_c|^2/4+(\delta+i\Gamma_{hg})^3}
{\big[|\Omega_c|^2/4-(\Delta+i\Gamma_{eg})(\delta+i\Gamma_{hg})\big]^3}. 
\end{eqnarray}
\end{widetext}
We substitute Eq. (\ref{e35}) into Eq. (\ref{e34}) and perform the inverse Fourier transformation. Then, we obtain
\begin{eqnarray}\label{e37}
\frac{\partial \mathcal{A}_f(z,t)}{f\partial z}&=&i(\beta_p+K_f) \mathcal{A}_f(z,t)-
(\beta'_p+K'_f)\frac{\partial \mathcal{A}_f(z,t)}{\partial t }
\nonumber\\&&\mbox{}
-i\frac{K''_f}{2}\frac{\partial^2 \mathcal{A}_f(z,t)}{\partial t^2 }.
\end{eqnarray}

In general, $K_f$, $K'_f$, and $K''_f$ are complex parameters. The imaginary and real parts of the parameter $K_f$, namely, the coefficients $\kappa_f=\mathrm{Im}(K_f)$ and $\theta_f=\mathrm{Re}(K_f)$, are, as already discussed above, the absorption and phase shift coefficients, respectively, for the guided probe light field.
We emphasize that the coefficients $K_f$, $K'_f$, and $K''_f$ are the propagation characteristics for the photon flux amplitude $\mathcal{A}_f(z,t)$.  

We analyze the case of exact one- and two-photon resonances, that is, the case where $\Delta=\delta=0$.
It is clear from the expression for $K_f$ in Eqs. (\ref{e36}) that, when $|\Omega_c|^2\gg \Gamma_{hg}\Gamma_{eg} $, the absorption coefficient $\kappa_f=\mathrm{Im}(K_f)$ and the phase shift coefficient $\theta_f=\mathrm{Re}(K_f)$ are small. These features are the signatures of EIT \cite{Harris review,Scully review,slow light review,most recent review,Scully,Agarwal book}. 
We note that the width of the corresponding transparency window is given by $\Delta\omega_{\mathrm{trans}}=1/\sqrt{L\,\mathrm{Im}( K''_f)}$. When the dephasing rate $\Gamma_{hg}$ is negligible, the expression for $K''_f$ in Eqs. (\ref{e36}) yields $\mathrm{Im}(K''_f)\propto \gamma_{eg}^{(f)}\Gamma_{eg}/\Lambda|\Omega_c|^{4}$, leading to \cite{Harris review,Scully review,slow light review,most recent review,Scully,Agarwal book}
\begin{equation}\label{e39}
\Delta\omega_{\mathrm{trans}}\propto \frac{|\Omega_c|^2}{\sqrt{\gamma_{eg}^{(f)}\Gamma_{eg}}}
\frac{1}{\sqrt{N}}.
\end{equation}
When the input probe pulse is long enough,
the group velocity $V_g^{(f)}$ of the probe field is determined by the equation 
${1}/{V_g^{(f)}}={1}/{v_g}+\textrm{Re}(K'_f)$.
When the dephasing rate $\Gamma_{hg}$ is negligible, the expression for $K'_f$ in Eqs. (\ref{e36}) yields $\mathrm{Re}(K'_f)\propto \gamma_{eg}^{(f)}/\Lambda|\Omega_c|^{2}$, leading to \cite{Harris review,Scully review,slow light review,most recent review,Scully,Agarwal book}
\begin{equation}\label{e41}
\frac{1}{V_g^{(f)}}-\frac{1}{v_g}\propto n_A \frac{\gamma_{eg}^{(f)}}{|\Omega_c|^{2}},
\end{equation}
where the atom-number density $n_A$ in the framework of the homogeneous-medium approximation is given by Eq.~\eqref{e19b}.

\begin{figure}[tbh]
\begin{center}
  \includegraphics{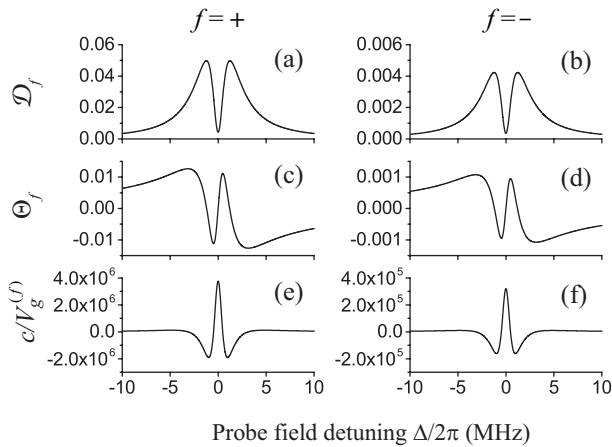}
\end{center}
\caption{
EIT of an $x$-polarized guided probe field propagating along an array of cesium atoms with a $\sigma$-type probe transition. The optical depth per atom $\mathcal{D}_f$ (upper row),
the phase shift per atom $\Theta_f$ (middle row), and the group-velocity reduction factor $c/V_g^{(f)}$ (lower row) are plotted as functions of the detuning $\Delta$. 
The working levels are $|e\rangle=|6P_{3/2},F'=4, M'=4\rangle$, $|g\rangle=|6S_{1/2},F=3, M=3\rangle$, and $|h\rangle=|6S_{1/2},F=4, M=4\rangle$. The lower-level decoherence rate is $\Gamma_{hg}=2\pi\times50$ kHz. The probe field propagates in the positive (left column) or negative (right column) direction of the fiber axis $z$. The fiber radius is $a=250$ nm. The distance from the atoms to the fiber surface is $r-a=200$ nm. The atom-number density is $n_A=1/\Lambda$ with $\Lambda=498.13$~nm. The control field is at exact resonance with the atomic transition $|e\rangle\leftrightarrow |h\rangle$ and is a plane wave propagating along the $x$ direction and polarized along the $y$ direction. The intensity of the control field is $I_c=1$ mW/cm$^2$ (the corresponding Rabi frequency is $\Omega_c=0.46\gamma_0$). 
} 
\label{fig2}
\end{figure}

\begin{figure}[tbh]
\begin{center}
  \includegraphics{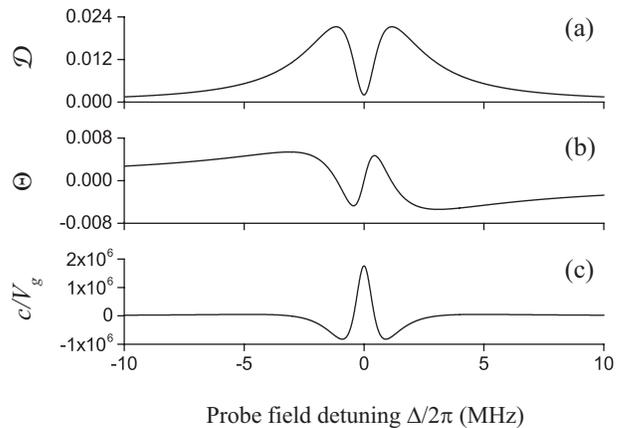}
\end{center}
\caption{
EIT of a $y$-polarized guided probe field propagating along an array of cesium atoms with a $\pi$-type probe transition. 
The optical depth per atom $\mathcal{D}$ (a), the phase shift per atom $\Theta$ (b), and the group-velocity reduction factor $c/V_g$ (c) are plotted as functions of the detuning $\Delta$. 
The working levels are $|e\rangle=|6P_{3/2},F'=4, M'=4\rangle$, $|g\rangle=|6S_{1/2},F=4, M=4\rangle$, and $|h\rangle=|6S_{1/2},F=3, M=3\rangle$. The lower-level decoherence rate is $\Gamma_{hg}=2\pi\times50$ kHz. The control field is at exact resonance with the atomic transition $|e\rangle\leftrightarrow |h\rangle$ and is a plane wave propagating along the $y$ direction with the counterclockwise circular polarization. The intensity of the control field is $I_c=1$ mW/cm$^2$ (the corresponding Rabi frequency is $\Omega_c=0.43\gamma_0$). Other parameters are as for Fig.~\ref{fig2}. 
The results do not depend on the propagation direction of the guided probe field. 
} 
\label{fig3}
\end{figure}

\begin{figure}[tbh]
\begin{center}
  \includegraphics{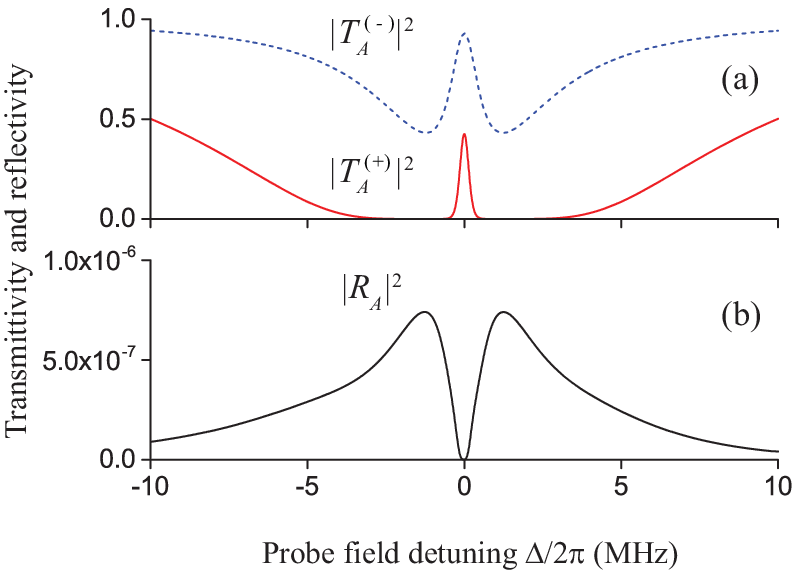}
\end{center}
\caption{(Color online)
Transmittivity $|T_A^{(f)}|^2$ (a) and reflectivity $|R_A|^2$ (b) of an $x$-polarized guided probe field as functions of the detuning $\Delta$ in the homogeneous-medium approximation. The atom-number density is $n_A=1/\Lambda$ with $\Lambda=498.13$~nm. The medium length is $L=(N-1)\Lambda$ with $N=200$. Other parameters are as for Fig.~\ref{fig2}. 
}
\label{fig4}
\end{figure}

\begin{figure}[tbh]
\begin{center}
  \includegraphics{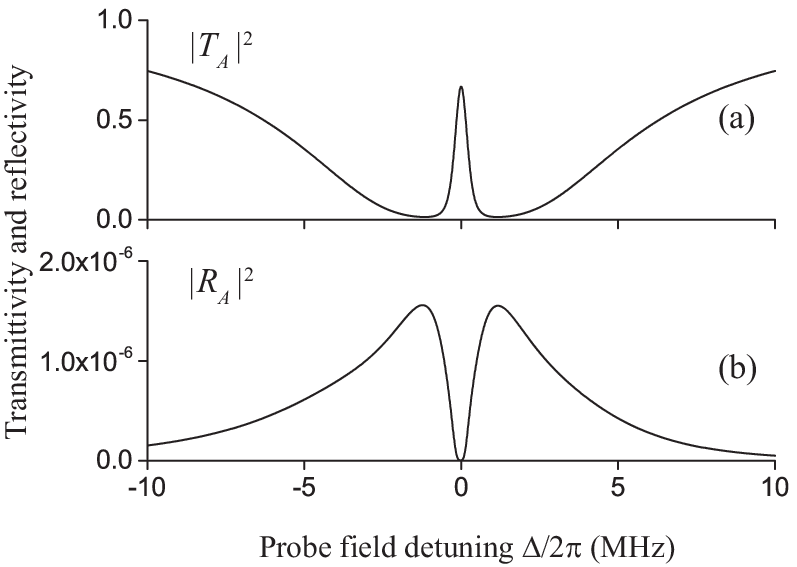}
\end{center}
\caption{
Transmittivity $|T_A|^2$ (a) and reflectivity $|R_A|^2$ (b) of a $y$-polarized guided probe field as functions of the detuning $\Delta$ in the homogeneous-medium approximation. The atom-number density is $n_A=1/\Lambda$ with $\Lambda=498.13$ nm. The medium length is $L=(N-1)\Lambda$ with $N=200$. Other parameters are as for Fig.~\ref{fig3}. 
}
\label{fig5}
\end{figure}

We calculate numerically the optical depth per atom $\mathcal{D}_f$, the phase shift per atom $\Theta_f$, the group-velocity reduction factor $c/V_g^{(f)}$, the transmittivity $|T_A^{(f)}|^2$, and the reflectivity $|R_A|^2$ as functions of the detuning $\Delta$ of the guided probe field. In the numerical calculations presented in this paper, we use, as already stated, the outermost Zeeman sublevels of the hyperfine levels $6P_{3/2}F'=4$, $6S_{1/2}F=3$, and $6S_{1/2}F=4$ of the $D_2$ line of atomic cesium, with the free-space wavelength $\lambda_0=852.35$ nm. In the calculations for the $x$-polarization scheme, we use the levels $|e\rangle=|6P_{3/2},F'=4, M'=4\rangle$, $|g\rangle=|6S_{1/2},F=3, M=3\rangle$, and $|h\rangle=|6S_{1/2},F=4, M=4\rangle$. In the calculations for the $y$-polarization scheme, we use the levels $|e\rangle=|6P_{3/2},F'=4, M'=4\rangle$, $|g\rangle=|6S_{1/2},F=4, M=4\rangle$, and $|h\rangle=|6S_{1/2},F=3, M=3\rangle$. The rate $\Gamma_{eg}$ of decay of the atomic optical-transition coherence is calculated by using the results of Ref. \cite{cesium decay}. The obtained value is $\Gamma_{eg}\simeq 2\pi\times 2.67$ MHz. The corresponding value of the linewidth of the upper level $|e\rangle$ is $2\Gamma_{eg}\simeq 2\pi\times 5.34$ MHz. This value is slightly larger than the literature value $\gamma_0\simeq 2\pi\times5.2$ MHz for the atomic natural linewidth \cite{Steck02,coolingbook}. The obtained increase of the atomic linewidth is caused by the presence of the nanofiber. The lower-level decoherence rate is assumed to be $\Gamma_{hg}=2\pi\times50$ kHz. This value is comparable to the experimental value of about $2\pi\times 32$ kHz, measured in the Vienna experiment \cite{Mitsch14a}. The control field is at exact resonance with the atomic transition $|e\rangle\leftrightarrow |h\rangle$. 
The fiber radius is $a=250$ nm and the distance from the atoms to the fiber surface is $r-a=200$ nm \cite{Vetsch10}.

We plot in Figs.~\ref{fig2} and \ref{fig4} the results of the calculations for scheme (a) of Fig.~\ref{fig1}, where the guided probe field is $x$ polarized. We show in Figs.~\ref{fig3} and \ref{fig5} the results for scheme (b) of Fig.~\ref{fig1}, where the guided probe field is $y$ polarized. For the calculations of Figs.~\ref{fig2}--\ref{fig5}, we use the one-dimensional atom-number density $n_A=1/\Lambda$, where $\Lambda=498.13$ nm. The chosen value of $\Lambda$ is one-half of the in-fiber wavelength of the red-detuned standing-wave guided light field used in the Vienna atom trap experiment \cite{Vetsch10}. Such a value of the array period is far from the Bragg resonance (the Bragg resonant array period is $\Lambda_{\mathrm{res}}=n\lambda_F/2$, where $n=1,2,\dots$ and $\lambda_F\equiv 2\pi/\beta_p\simeq 745.16$ nm for the probe field with the free-space atomic resonance wavelength $\lambda_p=\lambda_0=852.35$ nm).

Figures \ref{fig2} and \ref{fig3} show the familiar features of EIT: in the vicinity of the exact resonance, the optical depth $\mathcal{D}_f$ achieves a low minimum, the magnitude of the phase shift  $\Theta_f$ is small but the slope of the phase shift is steep, and consequently the reduction of the group velocity $V_g^{(f)}$ is large \cite{Harris review,Scully review,slow light review,most recent review,Scully,Agarwal book,vapor,ultracold,polariton}. These features occur in a region of frequency which is narrower than the atomic natural linewidth $\gamma_0$ and is called the EIT window. 

Figures~\ref{fig4}(a) and \ref{fig5}(a) show that the transmittivity $|T_A^{(f)}|^2$ has a narrow peak at $\Delta=0$. 
We observe from Figs. \ref{fig4}(b) and \ref{fig5}(b) that the reflectivity $|R_A|^2$ is very small and is slightly asymmetric with respect to $\Delta$. The asymmetry results from the fact that the relative phase $\beta_pL$ is not an integer multiple of $\pi$ for the parameters used. Due to this fact, the magnitude of the reflectivity $|R_A|^2$ depends on the sign of the detuning $\Delta$. 

When we compare the left and right columns of Fig.~\ref{fig2}, where the probe field polarization is $\xi_p=x$ and the propagation direction is $f=+$ for the left column and $f=-$ for the right column, and compare these two columns with Fig.~\ref{fig3}, where $\xi_p=y$ and $f=\pm$, we see that the optical depth per atom $\mathcal{D}_f$, the phase shift per atom $\Theta_f$, the group-velocity reduction factor $c/V_g^{(f)}$ are different in the three cases. Similarly, when we compare the solid red and dashed blue curves of Fig.~\ref{fig4}(a), where the probe field polarization is $\xi_p=x$ and the propagation direction is $f=+$ for the solid red curve and $f=-$ for the dashed blue curve, and compare these two curves with the curve of Fig.~\ref{fig5}(a), where $\xi_p=y$ and $f=\pm$, we observe that the transmittivity $|T_A^{(f)}|^2$ is also different in the three cases. The differences between the results for the three cases arise from the fact that the atom-field coupling characterized by the coefficient $G_f$ depends on the local polarization of the guided probe field at the position of the atom [see Eq. \eqref{e10}], that is, on the mode polarization $\xi_p=x,y$ and on the propagation direction $f$ in the case of $\xi_p=x$ [see Eqs. \eqref{e12} and \eqref{e11}]. Another reason is that different atomic transitions $|e\rangle\leftrightarrow|g\rangle$, which have different complex dipole matrix elements $\mathbf{d}_{eg}$, are used in the $x$- and $y$-polarization schemes (see Fig.~\ref{fig1}). 

Thus, in the case of the $x$-polarization scheme, the optical depth $\mathcal{D}_f$, the phase shift $\Theta_f$, the group velocity $V_g^{(f)}$, and the transmittivity $|T_A^{(f)}|^2$ substantially depend on the propagation direction $f$. The directionality of transmission of guided light through the array of atoms is a consequence of the existence of a longitudinal component of the guided light field as well as the ellipticity of both the field polarization and the atomic dipole vector \cite{AtomArray,Fam14}. We note that directional spontaneous emission into an optical nanofiber has been recently demonstrated experimentally for trapped atoms \cite{Mitsch14b} and for nanoparticles \cite{Petersen14}. An optical diode based on the chirality of guided photons has also been reported \cite{Sayrin15b}.

We observe from Figs.~\ref{fig2}(a), \ref{fig2}(b), and \ref{fig4}(a) for the $x$-polarization scheme and Figs.~\ref{fig3}(a) and \ref{fig5}(a) for the $y$-polarization scheme that,
at zero detuning, the absorption is not completely suppressed and, hence, the induced transparency is not 100\%. This is a consequence of the fact that a nonzero lower-level decoherence
rate $\Gamma_{hg}$ is included in our calculations. Without this decoherence, the induced transparency is 100\% at zero detuning. Comparison between Figs.~\ref{fig2}(a),  \ref{fig2}(b), and 
\ref{fig3}(a) and between Figs.~\ref{fig4}(a) and  \ref{fig5}(a) shows that, for a given lower-level decoherence, the residual absorption depends on the probe field polarization $\xi_p$
and on the propagation direction $f$ in the case of $\xi_p=x$. The reason is twofold: (1) the atom-field coupling depends on the local polarization of the field 
and (2) the different atomic transitions $|e\rangle\leftrightarrow|g\rangle$ are used in the different polarization schemes.

\subsection{Phase-matching approximation}
\label{subsec:phase-matching}

We now take into account the periodicity of the atomic array but still consider the array as a continuous one-dimensional medium. For this purpose, 
we rewrite expression \eqref{e9a} for the one-dimensional atom-number distribution $n_A$ in the Fourier series form
$n_A=(1/\Lambda)\sum_{n=-\infty}^\infty e^{2n\pi i z/\Lambda}$ for $0\le z\le L=(N-1)\Lambda$.
Then, Eq.~(\ref{e17}) for the coupling coefficients $\tilde{K}_{ff'}$ becomes 
\begin{equation}\label{e93}
\tilde{K}_{ff'}=K_{ff'}\sum_{n=-\infty}^\infty e^{2ni\beta_{\mathrm{lat}}z}.
\end{equation}
Here, the parameter $\beta_{\mathrm{lat}}=\pi/\Lambda$ is the wave number that characterizes the periodicity of the atom distribution in the array. The coupling coefficients $K_{ff'}$ are independent of $z$ and are given by Eq.~\eqref{e19}. 

We assume that there exists a positive integer number $n=1,2,\dots$ such that the propagation constant $\beta_p$ for the guided probe field is close to $n\beta_{\mathrm{lat}}$. This specific integer number $n$ indicates the dominant role of the corresponding harmonic in Eq.~\eqref{e93}. We note that the equality $\beta_p=n\beta_{\mathrm{lat}}$ means $\beta_p\Lambda=n\pi$, which is the geometric condition for the $n$th-order Bragg resonance.

We introduce the transformation $\mathcal{A}_{+}=a_{+}e^{in\beta_{\mathrm{lat}}z}$ and $\mathcal{A}_{-}=a_{-}e^{-in\beta_{\mathrm{lat}}z}$. When we insert Eq.~\eqref{e93} into Eqs.~\eqref{e16} and keep only the phase-matching terms, we obtain
\begin{eqnarray}\label{e94}
\left(\frac{\partial }{\partial z}-i\delta\beta_p\right)a_{+}&=&
iK_{++}a_{+}+iK_{+-}a_{-},\nonumber\\
\left(\frac{\partial }{\partial z}+i\delta\beta_p\right)a_{-}&=&
-iK_{-+}a_{+}-iK_{--}a_{-}.
\end{eqnarray}
Here, the factor $\delta\beta_p=\beta_p-n\beta_{\mathrm{lat}}$ characterizes the phase mismatch between the guided probe field and the $n$th harmonic of the atomic distribution. In Eqs.~\eqref{e94}, the terms $iK_{++}a_{+}$ and $-iK_{--}a_{-}$ describe the self coupling of the modes and originate from the zeroth-order harmonic in the Fourier expansion of $n_A$, while the terms $iK_{+-}a_{-}$ and $-iK_{-+}a_{+}$ describe the coupling between the forward and backward fields and originate from the $n$th-order harmonic in the Fourier expansion of $n_A$.
Similar to the results of the previous subsection, the reflection and transmission coefficients in the phase-matching approximation are found to be
\begin{eqnarray}\label{e95}
R_A&=&\frac{iK_{-+}\sin(UL)}{U\cos(UL)-i[\delta\beta_p+(K_{+}+K_{-})/2]\sin(UL)},
\nonumber\\
T_A^{(f)}&=&\frac{U\exp[if(K_{+}-K_{-})L/2]\exp(in\beta_{\mathrm{lat}}L)}{U\cos(UL)-i[\delta\beta_p+(K_{+}+K_{-})/2]\sin(UL)},\qquad
\end{eqnarray}
where
\begin{equation}\label{e96}
U=\sqrt{(\delta\beta_p)^2+\delta\beta_p(K_{+}+K_{-})+\frac{(K_{+}-K_{-})^2}{4}}.
\end{equation}

\begin{figure}[tbh]
\begin{center}
  \includegraphics{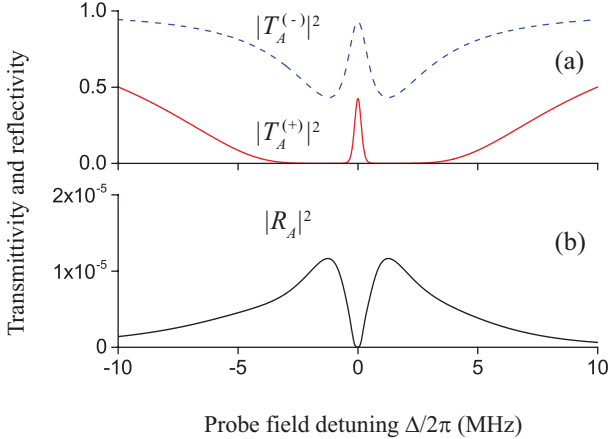}
\end{center}
\caption{(Color online)
Transmittivity $|T_A^{(f)}|^2$ (a) and reflectivity $|R_A|^2$ (b) of an $x$-polarized guided probe field as functions of the detuning $\Delta$ in the phase-matching approximation. The array period is $\Lambda=498.13$ nm and the array length is $L=(N-1)\Lambda$, where $N=200$. Other parameters are as for Fig.~\ref{fig2}. 
}
\label{fig6}
\end{figure}

\begin{figure}[tbh]
\begin{center}
  \includegraphics{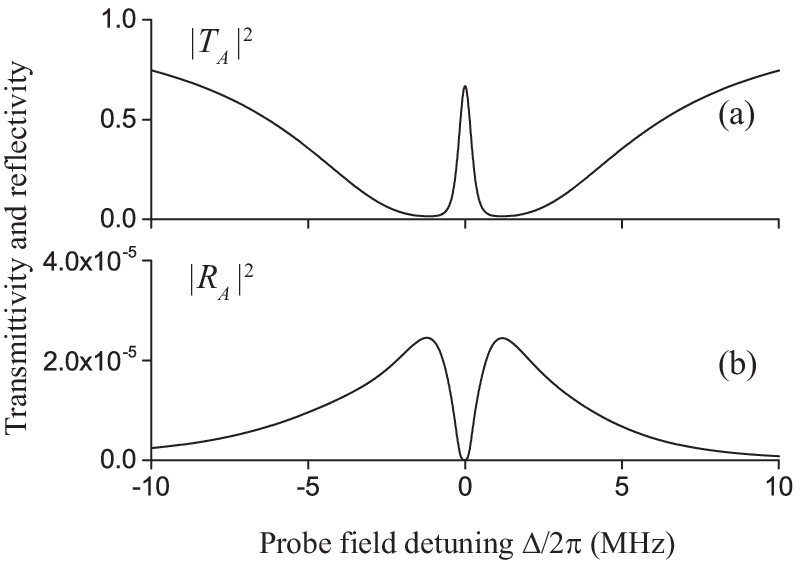}
\end{center}
\caption{
Transmittivity $|T_A|^2$ (a) and reflectivity $|R_A|^2$ (b) of a $y$-polarized guided probe field as functions of the detuning $\Delta$ in the phase-matching approximation. 
The array period is $\Lambda=498.13$ nm and the array length $L=(N-1)\Lambda$, where $N=200$. Other parameters are as for Fig.~\ref{fig3}. 
}
\label{fig7}
\end{figure}

The condition for the validity of the above results is $|\delta\beta_p|\ll\beta_p$. When the phase-mismatch factor $\delta\beta_p$ is sufficiently large that $|K_{ff'}|\ll|\delta\beta_p|$, 
Eqs.~\eqref{e95} reduce to
\begin{eqnarray}\label{e28a}
R_A&\simeq&\frac{K_{-+}}{2\delta\beta_p+K_{+}+K_{-}}\{\exp [i(2\delta\beta_p+K_{+}+K_{-})L]-1\},
\nonumber\\
T_A^{(f)}&\simeq&\exp [i(\beta_p+K_f) L].
\end{eqnarray}
Under the condition $|K_{-+}|\ll|\delta\beta_p|$, the reflectivity of the array is 
$|R_A|^2\simeq 0$, while
the transmittivity is $|T_A^{(f)}|^2\simeq\exp(-2\kappa_f L)$, with the absorption coefficient $\kappa_f$ being given by Eq.~\eqref{e32}.

We plot in Figs.~\ref{fig6} and \ref{fig7} the tuning dependences of the transmittivity $|T_A^{(f)}|^2$ and the reflectivity $|R_A|^2$, calculated from Eqs.~\eqref{e95} for the parameters
of Figs. \ref{fig4} and \ref{fig5}. We observe that Figs.~\ref{fig6}(a) and \ref{fig7}(a) are almost identical to Figs.~\ref{fig4}(a) and \ref{fig5}(a), respectively. 
However, the peak values of the curves in Figs.~\ref{fig6}(b) and \ref{fig7}(b) are one order of magnitude larger than that in Figs.~\ref{fig4}(b) and \ref{fig5}(b), respectively. 
The discrepancy is due to the fact that the periodicity of the atomic array is neglected in the homogeneous-medium approximation but is taken into account in the phase-matching approximation.

\section{Transfer matrix formalism for a discrete array}
\label{sec:array}

We now take into account not only the periodicity of the atomic position distribution $n_A=\sum_j\delta(z-z_j)$ but also the discrete nature of this distribution.
For this purpose, we use the transfer matrix formalism.

\subsection{Transfer matrix and input-output relation}
\label{subsec:input-output}

We first consider atom $j$ with the axial coordinate $z_j$. We introduce the notations $z_j^{\pm}=\lim_{\varepsilon\to 0_+} (z_j\pm\varepsilon)$ for the right- and left-hand-side limiting points. 
The fields $\mathcal{A}_{+}(z_j^{-})$ and $\mathcal{A}_{-}(z_j^{+})$
with the propagation directions $f=+$ and $f=-$, respectively, at the limiting points $z_j^{-}$ and $z_j^{+}$, respectively, can be interpreted as incoming fields with respect to the atom. Meanwhile, the fields $\mathcal{A}_{+}(z_j^{+})$ and $\mathcal{A}_{-}(z_j^{-})$ with the propagation directions $f=+$ and $f=-$, respectively, at the limiting points $z_j^{+}$ and $z_j^{-}$, respectively, can be considered as outgoing fields with respect to the atom. According to the causal principle, the atom interacts with the incoming fields $\mathcal{A}_{\pm}(z_j^{\mp})$
but not with the outgoing fields $\mathcal{A}_{\pm}(z_j^{\pm})$. 

We introduce the notations $\boldsymbol{\mathcal{A}}_L=\boldsymbol{\mathcal{A}}(z_1^{-})$ and $\boldsymbol{\mathcal{A}}_R=\boldsymbol{\mathcal{A}}(z_N^{+})$, where $\boldsymbol{\mathcal{A}}$ is the vector consisting of the components $\mathcal{A}_{f}$. 
When we integrate Eqs.~\eqref{e16} over an infinitely small interval $dz$ around the position $z_j$ of the atom and 
follow the procedures of Ref.~\cite{AtomArray}, we find the input-output relation
\begin{equation}\label{e56}
\boldsymbol{\mathcal{A}}_R=\mathbf{W}\boldsymbol{\mathcal{A}}_L,
\end{equation}
where
\begin{equation}\label{e57}
\mathbf{W}=\mathbf{T}^{N-1}\mathbf{M}
\end{equation}
is the transfer matrix for the whole atomic array, with
\begin{equation}\label{e58}
\mathbf{T}=\mathbf{M}\mathbf{F}
\end{equation}
being the transfer matrix for a single spatial period of the array.
The matrix $\mathbf{M}$ in Eqs.~\eqref{e57} and \eqref{e58} is the transfer matrix for a single atom. 
The matrix elements of this matrix are given as 
\begin{eqnarray}\label{x48a}
M_{++}&=&1-S_{++}+\frac{S_{+-}S_{-+}}{1+S_{--}},\nonumber\\
M_{--}&=&\frac{1}{1+S_{--}},\nonumber\\
M_{+-}&=&-\frac{S_{+-}}{1+S_{--}},\nonumber\\
M_{-+}&=&-\frac{S_{-+}}{1+S_{--}},
\end{eqnarray}
where
\begin{equation}\label{e50}
S_{ff'}=-if\Lambda K_{ff'}=-ifG_{f}^*G_{f'}\mathcal{F}.
\end{equation}
Note that we have $S_{+-}=-S_{-+}$ and $S_{++}S_{--}=S_{+-}S_{-+}$. 
The matrix $\mathbf{F}$ in Eq.~\eqref{e58} is the atom-free guided-field propagator. The elements of this matrix are given as
\begin{equation}\label{e55}
F_{ff'}=e^{if\beta_p\Lambda}\delta_{ff'}.
\end{equation}

It is easy to diagonalize the single-period transfer matrix $\mathbf{T}$. Using the result of this diagonalization, we find the following expressions for the elements of the transfer matrix $\mathbf{W}$ for the array:
\begin{eqnarray}
W_{++}&=&\mathcal{Z}^{N/2}\left\{\frac{M_{++}}{\sqrt{\mathcal{Z}}}\frac{\sin (N\zeta)}{\sin\zeta}
-e^{-i\beta_p \Lambda}\frac{\sin [(N-1)\zeta]}{\sin\zeta}\right\},
\nonumber\\
W_{--}&=&\mathcal{Z}^{N/2}\left\{\frac{M_{--}}{\sqrt{\mathcal{Z}}}\frac{\sin (N\zeta)}{\sin\zeta}
-e^{i\beta_p \Lambda}\frac{\sin [(N-1)\zeta]}{\sin\zeta}\right\},
\nonumber\\
W_{-+}&=&-W_{+-}=\mathcal{Z}^{N/2}\frac{M_{-+}}{\sqrt{\mathcal{Z}}}\frac{\sin (N\zeta)}{\sin\zeta}.
\label{x53}
\end{eqnarray}
Here, we have introduced the notations
\begin{equation}\label{e65} 
\mathcal{Z}=\det(\mathbf{M})=\frac{1-S_{++}}{1+S_{--}}=\frac{1+i|G_{+}|^2\mathcal{F}}{1+i|G_{-}|^2\mathcal{F}}
\end{equation} 
and
\begin{equation}\label{x54} 
\zeta=i\ln (D\pm i\sqrt{1-D^2}),
\end{equation} 
where
\begin{equation}\label{e66} 
D=\frac{1}{2\sqrt{\mathcal{Z}}}(M_{++}e^{i\beta_p \Lambda}+M_{--}e^{-i\beta_p \Lambda}).
\end{equation} 
Note that $\zeta$ is, in general, a complex parameter and satisfies the relations $\cos\zeta=D$ and $\sin\zeta=\mp\sqrt{1-D^2}$.

In the case of the $x$-polarization scheme, the elements of the transfer matrix $\mathbf{M}$ for a single atom are found to be
\begin{eqnarray}\label{e89}
M_{++}&=&1-\frac{S_r+S_z+2S_{rz}}{1-S_r-S_z+2S_{rz}},\nonumber\\
M_{--}&=&\frac{1}{1-S_r-S_z+2S_{rz}},\nonumber\\
M_{+-}&=&-M_{-+}=-\frac{S_r-S_z}{1-S_r-S_z+2S_{rz}}.
\end{eqnarray}
Here, we have introduced the notations
\begin{eqnarray}\label{e87}
S_r&=& -i\frac{\omega_pd_{eg}^2}{2\epsilon_0\hbar v_g}|e_r|^2\mathcal{F},\nonumber\\
S_z&=& -i\frac{\omega_pd_{eg}^2}{2\epsilon_0\hbar v_g}|e_z|^2\mathcal{F},\nonumber\\
S_{rz}&=& -i\frac{\omega_pd_{eg}^2}{2\epsilon_0\hbar v_g}(M_e-M_g)|e_r||e_z|\mathcal{F}.
\end{eqnarray}
We find $\mathcal{Z}\not=1$ in the case of the $x$-polarization scheme. 

In the case of the $y$-polarization scheme, the elements of the transfer matrix $\mathbf{M}$ for a single atom are found to be
\begin{eqnarray}\label{e90}
M_{++}&=&\frac{1-2S_\varphi}{1-S_\varphi},\nonumber\\
M_{--}&=&\frac{1}{1-S_\varphi},\nonumber\\
M_{+-}&=&-M_{-+}=-\frac{S_\varphi}{1-S_\varphi}.
\end{eqnarray}
Here, we have introduced the notations
\begin{equation}\label{e91}
S_\varphi= -i\frac{\omega_pd_{eg}^2}{\epsilon_0\hbar v_g}|e_\varphi|^2\mathcal{F}.
\end{equation}
We find $\mathcal{Z}=1$ in the case of the $y$-polarization scheme. 

\subsection{Reflection and transmission of probe light}
\label{subsec:reflection}

For the guided probe field propagating in the direction $f=+$, the reflection and transmission coefficients of a single atom are given by
$R^{(+)}=-M_{-+}/M_{--}$ and $T^{(+)}=\mathcal{Z}/M_{--}$, respectively.
The explicit expressions for these coefficients are
\begin{eqnarray}\label{x51}
R^{(+)}&=&S_{-+},\nonumber\\
T^{(+)}&=&1-S_{++}.
\end{eqnarray}
The corresponding reflection and transmission coefficients of the atomic array are given by
$R^{(+)}_N=-W_{-+}/W_{--}$ and $T^{(+)}_N=\mathcal{Z}^N/W_{--}$, respectively. 

For the guided probe field propagating in the direction $f=-$, the reflection and transmission coefficients of a single atom are given by
$R^{(-)}=M_{+-}/M_{--}$ and $T^{(-)}=1/M_{--}$, respectively.
The explicit expressions for these coefficients are
\begin{eqnarray}\label{e60}
R^{(-)}&=&-S_{+-},\nonumber\\
T^{(-)}&=&1+S_{--}.
\end{eqnarray}
The corresponding reflection and transmission coefficients of the atomic array are given by
$R^{(-)}_N=W_{+-}/W_{--}$ and $T^{(-)}_N=1/W_{--}$, respectively. 

It follows from the properties $S_{+-}=-S_{-+}$ and $W_{+-}=-W_{-+}$ that the single-atom reflection coefficient $R^{(f)}$ and the linear-array reflection coefficient $R^{(f)}_N$ do not depend on the field propagation direction $f$, that is, $R^{(+)}=R^{(-)}\equiv R$ and $R^{(+)}_N=R^{(-)}_N\equiv R_N$. However, the corresponding transmission coefficients $T^{(f)}$ and $T^{(f)}_N$ may depend on the propagation direction $f$. We find
\begin{eqnarray}\label{e67}
R_N&=&\frac{R\sin (N\zeta)}{\sin (N\zeta)-\sqrt{\mathcal{Z}}\,T^{(-)}e^{i\beta_p \Lambda}\sin [(N-1)\zeta]},\nonumber\\
T^{(f)}_N&=&\frac{\mathcal{Z}^{(1+fN)/2}\,T^{(-)}\sin\zeta}{\sin (N\zeta)-\sqrt{\mathcal{Z}}\,T^{(-)}e^{i\beta_p \Lambda}\sin [(N-1)\zeta]}.\qquad 
\end{eqnarray}
We can show that $R_N$ and $T^{(f)}_N$ satisfy the recurrence formulas
\begin{eqnarray}\label{x56}
R_{N+1}&=&R_N+\frac{T^{(+)}_N T^{(-)}_N R e^{2i\beta_p \Lambda}}{1-R_NRe^{2i\beta_p \Lambda}},
\nonumber\\
T^{(f)}_{N+1}&=&\frac{T^{(f)}_N T^{(f)}e^{i\beta_p \Lambda}}{1-R_NRe^{2i\beta_p \Lambda}},
\end{eqnarray}
which are in agreement with ray optics. It is clear that the reflection and transmission coefficients $R_N$ and $T^{(f)}_N$ 
for the atomic array depend on the polarization of the guided probe light field. We note that,
in the vicinity of a Bragg resonance, we can reduce Eqs.~\eqref{e67} to Eqs.~\eqref{e95}, which were obtained in the phase-matching approximation.

The phases of the transmission and reflection coefficients are given by $\Phi_T^{(f)}=\mathrm{Im}\,(\ln T^{(f)}_N)$ and $\Phi_R=\mathrm{Im}\,(\ln R_N)$, respectively. The group delays of the transmitted and reflected fields are given by $\tau_T^{(f)}=\Phi_T^{(f)\prime}$ and $\tau_R=\Phi_R^{\prime}$, respectively, where $\Phi'\equiv d\Phi/d\omega_p$. Thus, from the frequency dependences of the transmission and reflection coefficients we can calculate the group delays of the transmitted and reflected fields.

It follows from the expression for $T^{(f)}_N$ in Eqs.~\eqref{e67} that
\begin{equation}\label{e68}
\frac{T^{(+)}_N}{T^{(-)}_N}=\mathcal{Z}^{N}=\left(\frac{1+i|G_{+}|^2\mathcal{F}}{1+i|G_{-}|^2\mathcal{F}}\right)^{N}.
\end{equation}
In the case of the $x$-polarization scheme, we have $|G_{+}|\not=|G_{-}|$ and, hence, $\mathcal{Z}\not=1$. In this case, the single-atom  and linear-array transmission coefficients $T^{(f)}$ and $T^{(f)}_N$ depend on the field propagation direction $f$. In the case of the $y$-polarization scheme, we have $G_{+}=G_{-}$ and, hence, $\mathcal{Z}=1$. In this case, the single-atom transmission coefficient $T^{(f)}$, the linear-array transmission coefficient $T^{(f)}_N$, and the group delay $\tau_T^{(f)}$ are independent of the propagation direction $f$, that is, we have $T^{(+)}=T^{(-)}\equiv T$, $T^{(+)}_N=T^{(-)}_N\equiv T_N$, and $\tau_T^{(+)}=\tau_T^{(-)}\equiv \tau_T$.

The dependence of the single-atom transmission coefficient $T^{(f)}$ on the propagation direction $f=\pm$ in the case of $x$-polarized guided light is a consequence of the difference between the coupling coefficients $G_+$ and $G_-$ for the different propagation directions $+$ and $-$, respectively. The independence of the single-atom reflection coefficient $R$ from the field propagation direction $f$ is a consequence of the fact that $R$ is proportional to the product of both coupling coefficients $G_+$ and $G_-$. The dependence of the linear-array transmission coefficient $T^{(f)}_N$ on the propagation direction $f$ in the case of $x$-polarized guided light is a consequence of the difference between $T^{(+)}$ and $T^{(-)}$ for each atom in the array. The independence of the linear-array reflection coefficient $R_N$ from the field propagation direction $f$ is a consequence of the fact that all the atoms in the array considered here have the same reflection coefficient $R$.
We note that the above properties of reflection and transmission of guided light in the atomic array are different from that of light in a conventional Fabry-P\'{e}rot cavity formed by two mirrors, with the reflection coefficients $R_1^{(+)}=R_1^{(-)}\equiv R_1$ and $R_2^{(+)}=R_2^{(-)}\equiv R_2$ and the transmission coefficients $T_1^{(+)}=T_1^{(-)}\equiv T_1$ and 
$T_2^{(+)}=T_2^{(-)}\equiv T_2$. The transmission coefficient $T_{\mathrm{cav}}$ of such a cavity does not depend on the propagation direction. However, in the case where $R_1\not=R_2$ or $T_1\not=T_2$, the cavity reflection coefficient $R_{\mathrm{cav}}^{(f)}$ may depend on the propagation direction $f$.

\begin{figure}[tbh]
\begin{center}
  \includegraphics{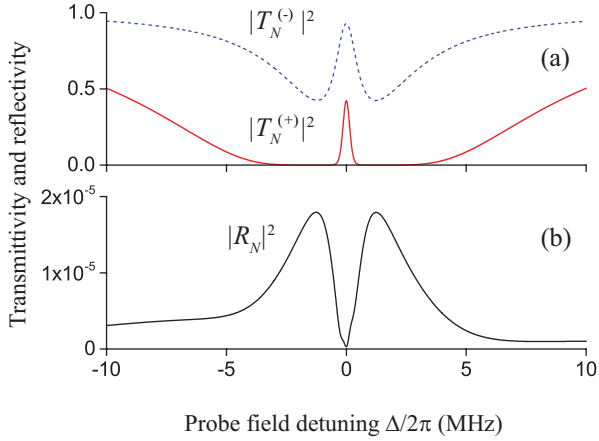}
\end{center}
\caption{(Color online)
Transmittivity $|T_N^{(f)}|^2$ (a) and reflectivity $|R_N|^2$ (b) of an $x$-polarized guided probe field as functions of the detuning $\Delta$.
The array period is $\Lambda=498.13$ nm, which is far from the Bragg resonance. The atom number is $N=200$. Other parameters are as for Fig.~\ref{fig2}. 
}
\label{fig8}
\end{figure}

\begin{figure}[tbh]
\begin{center}
  \includegraphics{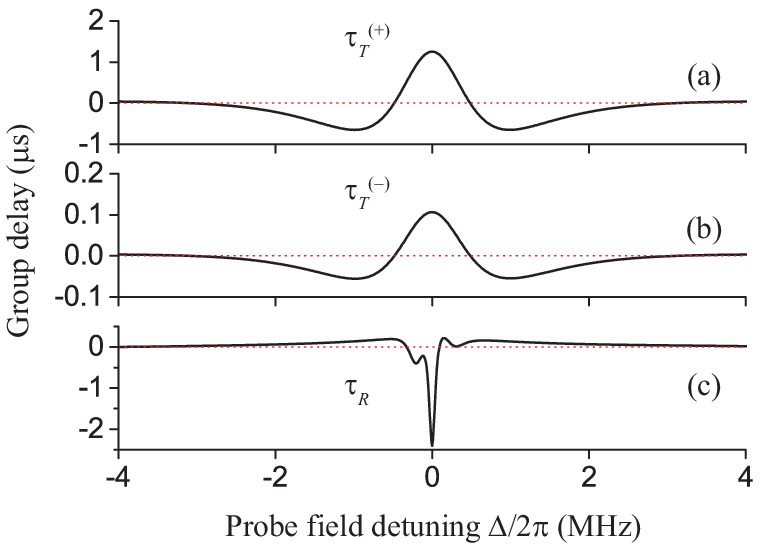}
\end{center}
\caption{(Color online)
Group delays $\tau_T^{(+)}$ (a), $\tau_T^{(-)}$ (b), and $\tau_R$ (c) of an $x$-polarized guided probe field as functions of the detuning $\Delta$. 
The array period is $\Lambda=498.13$ nm, which is far from the Bragg resonance. The atom number is $N=200$. 
Other parameters are as for Fig.~\ref{fig2}. The dotted red line is for the zero group delay and is a guide to the eye.
}
\label{fig9}
\end{figure}

\begin{figure}[tbh]
\begin{center}
  \includegraphics{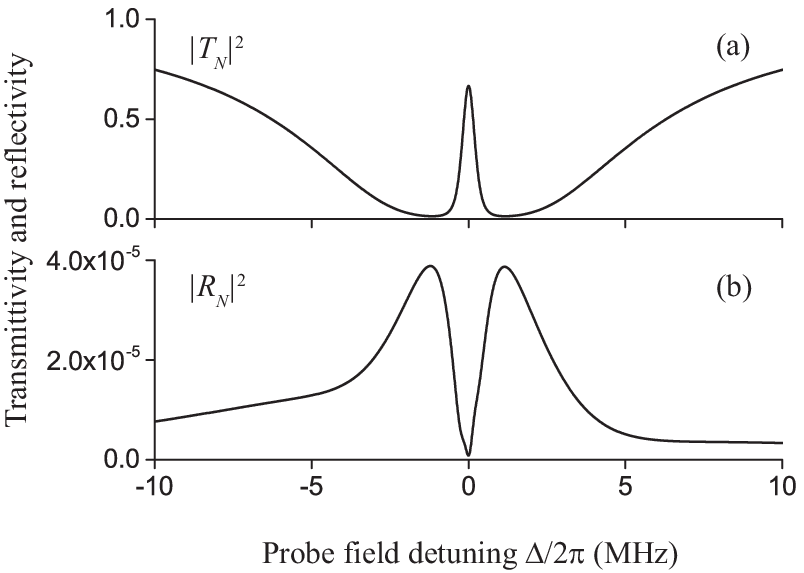}
\end{center}
\caption{
Transmittivity $|T_N|^2$ (a) and reflectivity $|R_N|^2$ (b) of a $y$-polarized guided probe field as functions of the detuning $\Delta$.
The array period is $\Lambda=498.13$ nm, which is far from the Bragg resonance. The atom number is $N=200$. Other parameters are as for Fig.~\ref{fig3}. 
}
\label{fig10}
\end{figure}

\begin{figure}[tbh]
\begin{center}
  \includegraphics{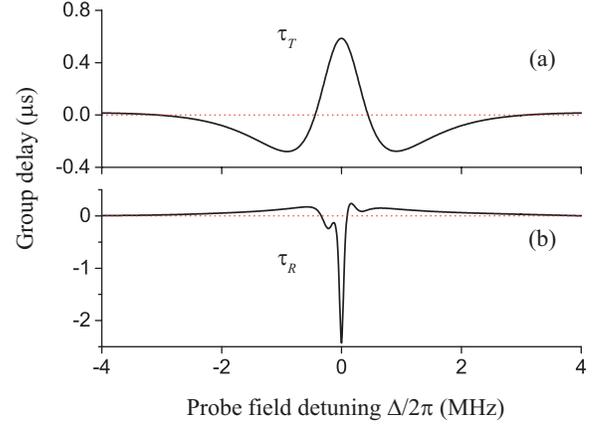}
\end{center}
\caption{(Color online)
Group delays $\tau_T$ (a) and $\tau_R$ (b) of a $y$-polarized guided probe field as functions of the detuning $\Delta$. 
The array period is $\Lambda=498.13$ nm, which is far from the Bragg resonance. The atom number is $N=200$. 
Other parameters are as for Fig.~\ref{fig3}. The dotted red line is for the zero group delay and is a guide to the eye.
}
\label{fig11}
\end{figure}

We calculate numerically the transmittivity $|T_N^{(f)}|^2$, the reflectivity $|R_N|^2$, the transmitted-field group delay $\tau_T^{(f)}$, and the reflected-field group delay $\tau_R$ as functions of the detuning $\Delta$ of the guided probe field. We plot in Figs.~\ref{fig8} and \ref{fig9} the results for the $x$-polarization scheme, and in Figs.~\ref{fig10} and \ref{fig11} the results for the $y$-polarization scheme. In the calculations for these figures, we used the value $\Lambda=498.13$ nm, which corresponds to the array period in the situation of the atom trap experiment \cite{Vetsch10}. This value of the array period is, as already noted in the previous section, far from the Bragg resonance. We observe that Figs.~\ref{fig8}(a) and \ref{fig10}(a) are almost identical to Figs.~\ref{fig4}(a) and \ref{fig5}(a), respectively. This means that the homogeneous-medium approximation is a very good approximation for the calculations of the transmittivity in the case where the array period is far from the Bragg resonance. Figures \ref{fig8}(b) and \ref{fig10}(b) show that the discrete-array reflectivity $|R_N|^2$, like the homogeneous-medium reflectivity $|R_A|^2$ in Figs. \ref{fig4}(b) and \ref{fig5}(b), is very small and is slightly asymmetric with respect to $\Delta$. However, the maximum magnitude of discrete-array reflectivity $|R_N|^2$ is about twenty times larger than that of the homogeneous-medium reflectivity $|R_A|^2$. Thus, the homogeneous-medium approximation is not reliable in evaluating the reflectivity. The discrepancy is due to the fact that the reflected field results from the interference between the fields reflected from different atoms in the array. The periodicity of the array greatly affects the interference but is neglected in the homogeneous-medium approximation. Comparison between Figs.~\ref{fig4}(b), \ref{fig6}(b), and \ref{fig8}(b) and between Figs.~\ref{fig5}(b), \ref{fig7}(b), and \ref{fig10}(b) shows that the reflectivity calculated in the phase-matching approximation is closer to the rigorous result $|R_N|^2$ than that calculated in the homogeneous-medium approximation.

Figures \ref{fig9}(a), \ref{fig9}(b), and \ref{fig11}(a) show that the group delay $\tau_T^{(f)}$ of the transmitted field has a positive peak at $\Delta=0$. This result means that the transmitted field is slow light in the vicinity of the atomic resonance. Close inspection shows that the results for the group delay $\tau_T^{(f)}$ plotted in Figs. \ref{fig9}(a), \ref{fig9}(b), and \ref{fig11}(a) are in full agreement with the results for the group-velocity reduction factor $c/V_g^{(f)}$ plotted in Figs.~\ref{fig2}(e), \ref{fig2}(f), and \ref{fig3}(c), respectively. According to Figs. \ref{fig9}(c) and \ref{fig11}(b), the group delay $\tau_R$ of the reflected field has a negative-valued dip at $\Delta=0$. This result means that the reflected field is fast light in the vicinity of the atomic resonance
\cite{Scully,Agarwal book,Boyd09,Wang00}. Comparison between the solid red and dashed blue curves of Fig.~\ref{fig8}(a) and between Figs.~\ref{fig9}(a) and \ref{fig9}(b) shows that the transmittivity $|T_N^{(f)}|^2$ and the corresponding group delay $\tau_T^{(f)}$ depend on the propagation direction $f$ in the case of the $x$-polarization scheme. This dependence is a consequence of the directional dependence of the coupling between the $x$-polarized guided probe field and
the atomic transition $|e\rangle\leftrightarrow|g\rangle$, which is of the $\sigma$ type with respect to the $y$ axis in the case considered [see Eq.~\eqref{e12}].
We observe from Figs.~\ref{fig9} and \ref{fig11} that the dip of the reflected-field group delay $\tau_R$ has a narrower width and a larger magnitude than the peak of the transmitted-field group delay $\tau_T^{(f)}$. The difference is due to the fact that the transmitted field is determined by the interference between the incident field and the single-atom scattered fields while the reflected field is determined by the interference between the fields reflected from different atoms.

\subsection{Time evolution of guided probe field pulses}
\label{subsec:time}

We now consider the time evolution of guided field pulses propagating along the atomic array. We assume that the guided input pulse is incident onto the atomic array in the $f$ direction.
The amplitudes $\mathcal{A}_{\mathrm{in}}(t)$ and $\tilde{\mathcal{A}}_{\mathrm{in}}(\omega)$ of the incident field in the time and frequency domains, respectively, are related to each other by the 
Fourier transformation 
$\tilde{\mathcal{A}}_{\mathrm{in}}(\omega)=(2\pi)^{-1/2}\int_{-\infty}^{\infty}\mathcal{A}_{\mathrm{in}}(t)e^{i\omega t} dt$.
The time-dependent transmitted field is
\begin{equation}\label{e106}
\mathcal{A}_{\mathrm{out}}(t)=\frac{1}{\sqrt{2\pi}} \int_{-\infty}^{\infty}T^{(f)}_N(\omega)\tilde{\mathcal{A}}_{\mathrm{in}}(\omega)e^{-i\omega t} d\omega,
\end{equation}
where $T^{(f)}_N(\omega)=T^{(f)}_N|_{\omega_p=\omega}$.
The time-dependent reflected field is
\begin{equation}\label{e107}
\mathcal{A}_{\mathrm{ref}}(t)=\frac{1}{\sqrt{2\pi}} \int_{-\infty}^{\infty}R_N(\omega)\tilde{\mathcal{A}}_{\mathrm{in}}(\omega)e^{-i\omega t} d\omega,
\end{equation}
where $R_N(\omega)=R_N|_{\omega_p=\omega}$.

\begin{figure}[tbh]
\begin{center}
  \includegraphics{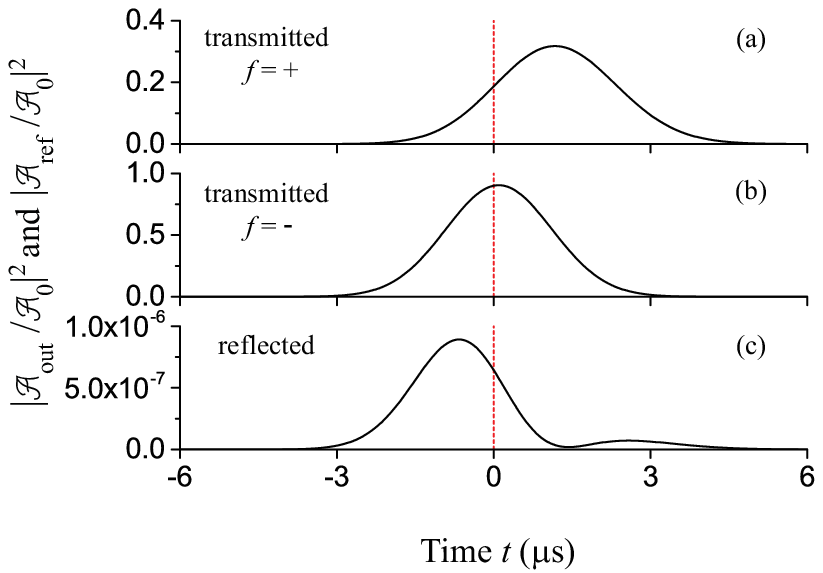}
\end{center}
\caption{(Color online)
Time dependences of the normalized intensities $|\mathcal{A}_{\mathrm{out,ref}}/\mathcal{A}_0|^2$ of the transmitted and reflected pulses in the case of the $x$-polarization scheme. The input pulse is of the Gaussian form, with the central frequency $\omega_p=\omega_0$, the pulse length $\tau_0=2$ $\mu$s, and the peak time $t=0$. The number of atoms is $N=200$. The array period is $\Lambda=498.13$ nm. Other parameters are as for Fig.~\ref{fig2}. The dotted red line is for the input pulse peak time $t=0$ and is a guide to the eye.
}
\label{fig12}
\end{figure}

\begin{figure}[tbh]
\begin{center}
  \includegraphics{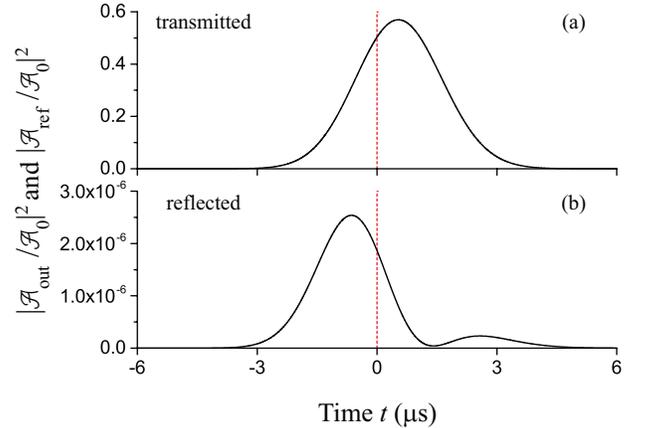}
\end{center}
\caption{(Color online)
Time dependences of the normalized intensities $|\mathcal{A}_{\mathrm{out,ref}}/\mathcal{A}_0|^2$ of the transmitted and reflected pulses in the case of the $y$-polarization scheme. The input pulse is of the Gaussian form, with the central frequency $\omega_p=\omega_0$, the pulse length $\tau_0=2$ $\mu$s, and the peak time $t=0$. The number of atoms is $N=200$. The array period is $\Lambda=498.13$ nm. Other parameters are as for Fig.~\ref{fig3}. The dotted red line is for the input pulse peak time $t=0$ and is a guide to the eye.
}
\label{fig13}
\end{figure}

We plot in Figs.~\ref{fig12} and \ref{fig13} the time dependences of the normalized intensities 
$|\mathcal{A}_{\mathrm{out,ref}}/\mathcal{A}_0|^2$ of the transmitted/reflected pulses in the cases of the $x$- and $y$-polarization schemes, respectively. In the numerical calculations for these figures, we used an input pulse of the Gaussian form $\mathcal{A}_{\mathrm{in}}(t)=\mathcal{A}_0e^{-t^2/\tau_0^2}e^{-i\omega_pt}$, where $\tau_0$ is the initial characteristic pulse length. The Fourier transform of the input pulse is $\tilde{\mathcal{A}}_{\mathrm{in}}(\omega)=(\mathcal{A}_0\tau_0/\sqrt2)e^{-\tau_0^2(\omega-\omega_p)^2/4}$. We observe that, when the probe pulse is incident onto the array in the direction $f=+$ in the case of the $x$-polarization scheme [see Fig.~\ref{fig12}(a)] and when the probe pulse is incident onto the array in an arbitrary direction $f=\pm$ in the case of the $y$-polarization scheme [see Fig.~\ref{fig13}(a)], the transmitted pulse is weakened and slowed down significantly by the atomic array. For $N=200$ atoms in the array with the length $L=(N-1)\Lambda\simeq 100$ $\mu$m, we obtain the group delays of about $1.17$ $\mu$s in the case of Fig.~\ref{fig12}(a) and $0.53$ $\mu$s in the case of Fig.~\ref{fig13}(a). These group delays correspond to the group-velocity reduction factors $c/V_g\simeq 3.5\times 10^6$ and $\simeq 1.6\times 10^6$ in the cases of Figs.~\ref{fig12}(a) and \ref{fig13}(a), respectively. Figure \ref{fig12}(b) shows that, when the probe pulse is $x$ polarized and incident onto the array in the direction $f=-$, the intensity reduction and the group delay of the transmitted pulse are not significant. According to Figs. \ref{fig12}(c) and \ref{fig13}(b), the reflected pulse has a negative group delay (fast light) \cite{Scully,Agarwal book,Boyd09,Wang00}. However, the intensity of this fast light is negligible.

\subsection{Bragg resonance}
\label{subsec:Bragg}

We examine the transmission and reflection of the guided probe field under the EIT condition in the specific case where the geometric Bragg resonance condition is satisfied, that is, $\beta_p\Lambda=n\pi$, with $n=1,2,\dots$ being the order of the Bragg resonance. This condition involves the frequency $\omega_p$ of the guided light field. Therefore, when we vary $\omega_p$ in an interval $\Delta\omega_p$ around a Bragg resonance frequency, the Bragg resonance condition $\beta_p\Lambda=n\pi$ will be broken. However, if the frequency variation interval $\Delta\omega_p$ is small as compared to the optical frequency $\omega_p$, the effect of the deviation from the Bragg resonance can be neglected. 

\subsubsection{Dependence on the probe field detuning}

\begin{figure}[tbh]
\begin{center}
  \includegraphics{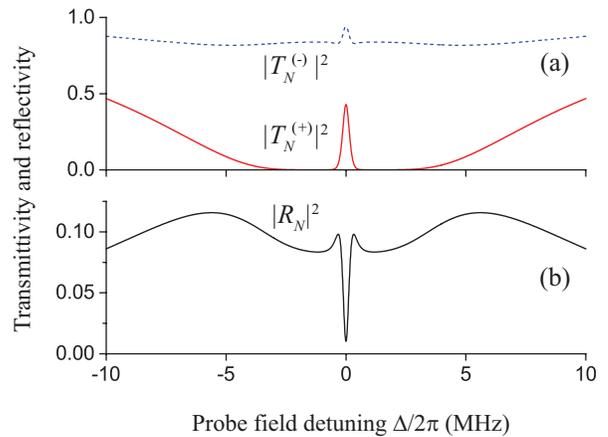}
\end{center}
\caption{(Color online)
Transmittivity $|T_N^{(f)}|^2$ (a) and reflectivity $|R_N|^2$ (b) of the atomic array in the $x$-polarization scheme as functions of the detuning $\Delta$.
The period of the array is $\Lambda=745.16$ nm, which satisfies the second-order Bragg resonance condition. 
The number of atoms in the array is $N=200$. Other parameters are as for Fig.~\ref{fig2}.
}
\label{fig14}
\end{figure}

\begin{figure}[tbh]
\begin{center}
  \includegraphics{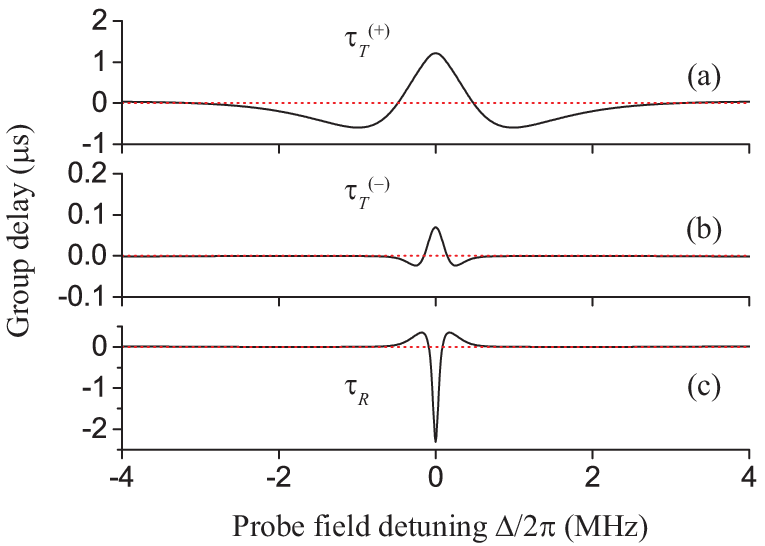}
\end{center}
\caption{(Color online)
Group delays $\tau_T^{(+)}$ (a), $\tau_T^{(-)}$ (b), and $\tau_R$ (c) in the $x$-polarization scheme as functions of the detuning $\Delta$. 
The period of the array is $\Lambda=745.16$ nm, which satisfies the second-order Bragg resonance condition. 
The atom number is $N=200$. Other parameters are as for Fig.~\ref{fig2}. The dotted red line is for the zero group delay and is a guide to the eye.
}
\label{fig15}
\end{figure}

\begin{figure}[tbh]
\begin{center}
  \includegraphics{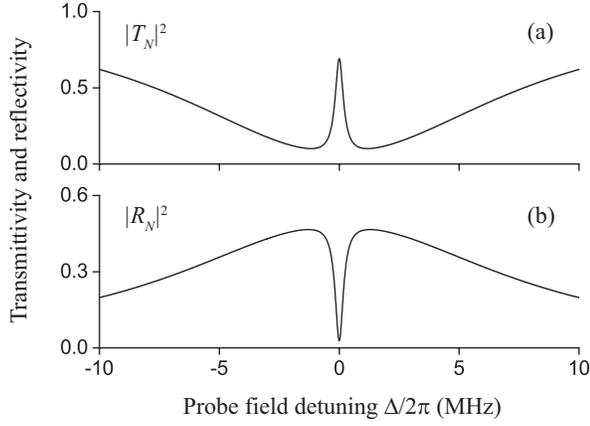}
\end{center}
\caption{
Transmittivity $|T_N|^2$ (a) and reflectivity $|R_N|^2$ (b) of the atomic array in the $y$-polarization scheme as functions of the detuning $\Delta$.
The period of the array is $\Lambda=745.16$ nm, which satisfies the second-order Bragg resonance condition. 
The atom number is $N=200$. Other parameters are as for Fig.~\ref{fig3}. 
}
\label{fig16}
\end{figure}

\begin{figure}[tbh]
\begin{center}
  \includegraphics{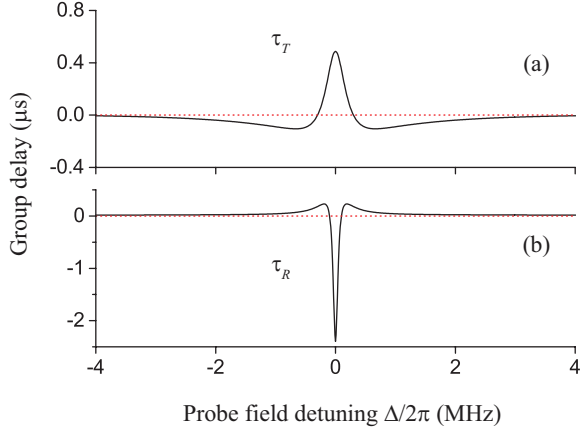}
\end{center}
\caption{(Color online)
Group delays $\tau_T$ (a) and $\tau_R$ (b) in the $y$-polarization scheme as functions of the detuning $\Delta$. 
The period of the array is $\Lambda=745.16$ nm, which satisfies the second-order Bragg resonance condition. 
The atom number is $N=200$. Other parameters are as for Fig.~\ref{fig3}. 
The dotted red line is for the zero group delay and is a guide to the eye.
}
\label{fig17}
\end{figure}

We plot in Figs.~\ref{fig14}--\ref{fig17} the transmittivity $|T_N^{(f)}|^2$, the reflectivity $|R_N|^2$, the transmitted-field group delay $\tau_T^{(f)}$, and the reflected-field group delay $\tau_R$ as functions of the detuning $\Delta$ of the guided probe field in the case where the array period is $\Lambda=745.16$ nm. This value of the array period satisfies the second-order Bragg resonance condition $\Lambda=\Lambda_{\mathrm{res}}=n\lambda_F/2$, where $n=2$ and $\lambda_F\equiv 2\pi/\beta_p\simeq 745.16$ nm for the probe field with the atomic resonance wavelength $\lambda_p=\lambda_0=852.35$ nm. 
We avoid the first-order Bragg resonance with the aim of minimizing the effects of the direct dipole-dipole interaction between the atoms.
It is worth mentioning here that, in the framework of our treatment, the magnitudes of the reflectivity $|R_N|^2$ and transmittivity $|T_N|^2$ of the atomic array for the guided fields do not depend on the order of the Bragg resonance.

Figures \ref{fig14} and \ref{fig15} show the numerical results for the $x$-polarization scheme. Comparison between Fig.~\ref{fig14}, where the array period $\Lambda$ is in the Bragg resonance, and Fig.~\ref{fig8}, where $\Lambda$ is far from the Bragg resonance, shows that the positive-direction ($f=+$) transmittivity $|T_N^{(+)}|^2$ is almost the same in the two cases [see the solid red lines in Figs.~\ref{fig14}(a) and \ref{fig8}(a)]. Meanwhile, the negative-direction ($f=-$) transmittivity $|T_N^{(-)}|^2$ in the Bragg-resonance case is higher than that in the far-off-Bragg-resonance case, except for a very narrow frequency region around $\Delta=0$, where it is the same [see the dashed blue lines in Figs.~\ref{fig14}(a) and \ref{fig8}(a)]. The reflectivity $|R_N|^2$ in the Bragg-resonance case [see Fig.~\ref{fig14}(b)] is dramatically higher than that in the far-off-Bragg-resonance case [see Fig.~\ref{fig8}(b)]. The dependence of $|R_N|^2$ on $\Delta$ is almost symmetric in the case of the Bragg resonance [see Fig.~\ref{fig14}(b)]. Comparison between Figs.~\ref{fig15}(a) and  \ref{fig9}(a) shows that the positive-direction transmitted-field group delay $\tau_T^{(+)}$ is almost the same in the two cases. 
Comparison between Figs.~\ref{fig15}(b) and \ref{fig9}(b) and between Figs.~\ref{fig15}(c) and \ref{fig9}(c) 
shows that the Bragg resonance modifies the values of the negative-direction transmitted-field group delay $\tau_T^{(-)}$ and the reflected-field group delay $\tau_R$. However, the changes are not dramatic. 

Figures \ref{fig16} and \ref{fig17} show the results of the numerical calculations for the $y$-polarization scheme. Comparison between Figs.~\ref{fig16}(a) and \ref{fig10}(a) and between Figs.~\ref{fig17} and \ref{fig11} shows that the effects of the Bragg resonance on the transmittivity $|T_N|^2$, the transmitted-field group delay $\tau_T$, and the reflected-field group delay $\tau_R$ are noticeable but not dramatic. Comparison between Figs.~\ref{fig16}(b) and \ref{fig10}(b) shows that, due to the Bragg resonance condition, the reflectivity $|R_N|^2$ is dramatically increased.

\subsubsection{Dependence on the atom number}

\begin{figure}[tbh]
\begin{center}
  \includegraphics{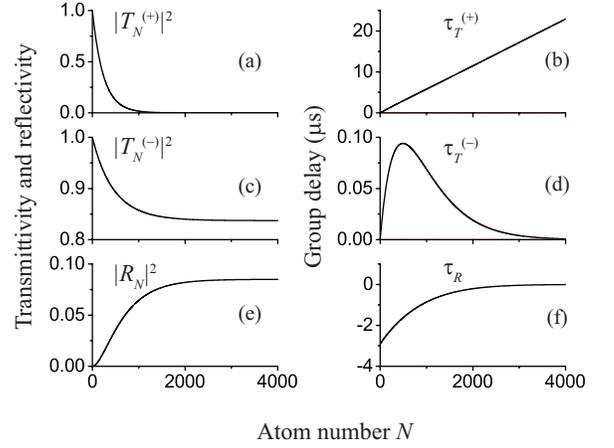}
\end{center}
\caption{
Dependences of the transmittivities $|T_N^{(+)}|^2$ and $|T_N^{(-)}|^2$, the reflectivity $|R_N|^2$, 
and the group delays $\tau_T^{(+)}$, $\tau_T^{(-)}$, and $\tau_R$ on the atom number $N$ in the case of the $x$-polarization scheme. The period of the array is $\Lambda=745.16$ nm, which satisfies the second-order Bragg resonance condition. 
The detuning of the probe field is $\Delta=0$. Other parameters are as for Fig.~\ref{fig2}.
}
\label{fig18}
\end{figure}

\begin{figure}[tbh]
\begin{center}
  \includegraphics{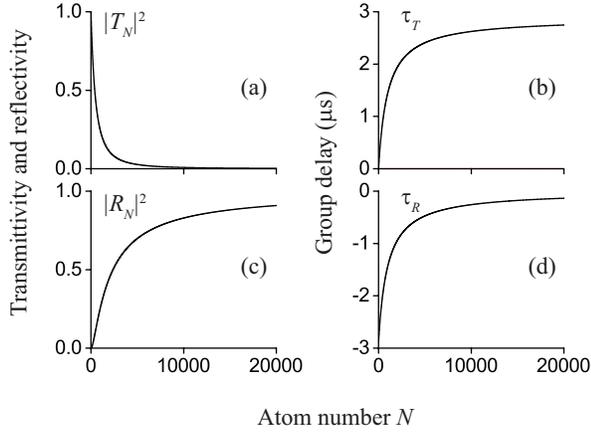}
\end{center}
\caption{
Dependences of the transmittivity $|T_N|^2$, the reflectivity $|R_N|^2$, and the group delays $\tau_T$ and $\tau_R$ on the atom number $N$ in the case of the $y$-polarization scheme. The period of the array is $\Lambda=745.16$ nm, which satisfies the second-order Bragg resonance condition. The detuning of the probe field is $\Delta=0$. Other parameters are as for Fig.~\ref{fig3}.
}
\label{fig19}
\end{figure}

We plot in Figs.~\ref{fig18} and \ref{fig19} the dependences of the transmittivity $|T_N^{(f)}|^2$, the reflectivity $|R_N|^2$, the transmitted-field group delay $\tau_T^{(f)}$, and the reflected-field group delay $\tau_R$ on the atom number $N$ for the array period $\Lambda=745.16$ nm, which satisfies the second-order Bragg resonance condition. We assume the atomic resonance $\Delta=0$ in these calculations. 

Figure \ref{fig18} shows the numerical results for the $x$-polarization scheme. We observe from the left column of Fig.~\ref{fig18} that, when $N$ increases, the positive-direction transmittivity $|T_N^{(+)}|^2$ decreases to zero, the negative-direction transmittivity $|T_N^{(-)}|^2$ decreases to a nonzero limiting value $|T_\infty^{(-)}|^2\simeq 0.84$ [see Eq.~\eqref{e76b} in the next part], and the reflectivity $|R_N|^2$ increases to a limiting value $|R_\infty|^2\simeq 0.08$, which is smaller than unity [see Eq.~\eqref{x60} in the next part]. Figures \ref{fig18}(b) and \ref{fig18}(d) show that, when $N$ increases, the positive-direction transmitted-field group delay $\tau_T^{(+)}$ increases almost linearly and hence can achieve an arbitrarily large value, while the negative-direction transmitted-field group delay $\tau_T^{(-)}$ increases to about $0.1$ $\mu$s and then decreases in the range $N\leq 4000$ of the figure. Additional calculations for $N>4000$ show that $\tau_T^{(-)}$ decreases to about 24 ps at $N\simeq 7400$ and then starts to slowly increase. We observe from Fig.~\ref{fig18}(f) that the reflected-field group delay $\tau_R$ is negative and its absolute value $|\tau_R|$ decreases with increasing $N$ in the range $N\leq 4000$ of the figure. Additional calculations for $N>4000$ show that $\tau_R$ becomes positive at $N\simeq 8440$ and then tends to increase very slowly. 

Figure \ref{fig19} shows the numerical results for the $y$-polarization scheme. We observe from the figure that, when $N$ increases, the reflectivity $|R_N|^2$ increases and approaches unity [see Eq.~\eqref{x66aa} in the next part], the transmitted-field group delay $\tau_T$ approaches a limiting value of about 3 $\mu$s, and the reflected-field group delay $\tau_R$ is negative and approaches zero.

\subsubsection{Limiting values}

We derive analytical expressions for the transmittivity and reflectivity of the atomic array in the limit of large $N$.

We first analyze the case of the $x$-polarization scheme under the Bragg resonance condition. In this case,
in the limit $N\to\infty$, we obtain the reflection coefficient 
\begin{equation}\label{x60}
R_N\to R_\infty= -\frac{|e_r|-|e_z|}{|e_r|+|e_z|}
\end{equation}
and the transmission coefficients
\begin{subequations}\label{e76} 
\begin{eqnarray}
|T_N^{(f_0)}|&\to& 0, \label{e76a}\\
|T_N^{(-f_0)}|&\to& |T_\infty^{(-f_0)}|=\frac{4|e_r||e_z|}{(|e_r|+|e_z|)^2}\not=0, \label{e76b}
\end{eqnarray}
\end{subequations} 
where $f_0=\mathrm{sign}(M_e-M_g)$.

It is clear that the limiting value $R_\infty$ is determined by the guided-mode profile functions $e_r$ and $e_z$ only.
Since $|e_r|> |e_z|>0$, we have $|R_\infty|^2<1$, that is, the limiting value $|R_\infty|^2$ of the reflectivity for the $x$-polarized guided fields is strictly smaller than unity. It is interesting to note that Eq.~\eqref{x60} coincides with the result of Ref.~\cite{Fam14} for the case where the initial state of the atoms is an incoherent mixture of the Zeeman sublevels of the ground state.

It is surprising that, unlike the transmittivity $|T_N^{(f_0)}|^2$ for the propagation direction $f_0$, the transmittivity $|T_N^{(-f_0)}|^2$ for the propagation direction $-f_0$ does not reduce to zero with increasing atom number $N$. Due to this feature, the atomic array can act as an optical diode even in the limit $N\to\infty$ \cite{Sayrin15b}. 
The property $|T_N^{(-f_0)}| \to |T_\infty^{(-f_0)}|\not=0$ in the limit $N\to\infty$ is not related to the EIT effect. Indeed, this property occurs even 
in the case where there is no control field ($\Omega_c=0$). This property is a consequence of the Bragg resonance condition and 
the difference between the coupling coefficients $|G_{+}|$ and $|G_{-}|$ for the guided modes with the opposite propagation directions $f=+$ and $f=-$, respectively. 
The difference between $|G_{+}|$ and $|G_{-}|$ is related to the existence of the longitudinal component $e_z$ of the $x$-polarized guided field \cite{Fam14}.

In order to get insight into the above result, we consider the change from the case of $N$ atoms to the case of $N+1$ atoms by adding an atom to the array of $N$ atoms under the Bragg resonance condition $\beta_p \Lambda=n\pi$. We use the recurrence relations \eqref{x56} to describe the change. 
In the limit $N\to\infty$, we have
$R_N\to R_\infty$ and $T_N^{(f_0)}\to 0$. 
We find that a nonzero asymptotic solution $T_N^{(-f_0)}\to (-1)^{(N+1)n} T_\infty^{(-f_0)}\not=0$ exists if
\begin{equation}\label{e80}
\frac{T^{(-f_0)}}{1-R_\infty R}=1,
\end{equation}
that is, if $R_\infty =(1-T^{(-f_0)})/R$, in agreement with Eq.~\eqref{x60}. 
The inverse of the denominator $1-R_\infty R$ in the expression on the left-hand side of condition \eqref{e80} characterizes the enhancement due to multiple reflections between 
the initial array and the added atom. Condition \eqref{e80} requires that the enhancement of transmission caused by multiple reflections compensates the reduction caused by a single pass through the added atom. 
Thus, in the limit of large $N$, the reflection coefficient $R_N$ of the array approaches an appropriate value $R_\infty$, which satisfies condition \eqref{e80}, and hence
the transmittivity $|T_N^{(-f_0)}|^2$ for the propagation direction $-f_0=-\mathrm{sign}(M_e-M_g)$ tends to a nonzero value given by Eq.~\eqref{e76b}.

We now analyze the case of the $y$-polarization scheme under the Bragg resonance condition. In this case, we find 
\begin{eqnarray}\label{e78}
R_N&=&\frac{NR}{1-(N-1)R},\nonumber\\
T_N&=&(-1)^{(N+1)n}\frac{T}{1-(N-1)R}.
\end{eqnarray}
In the limit $N\to\infty$, we have 
\begin{subequations}\label{x66a}
\begin{eqnarray}
R_N&\to&-1,\label{x66aa}\\
T_N&\to&0,\label{x66ab}
\end{eqnarray}
\end{subequations} 
that is, $|R_N|^2\to1$ and $|T_N|^2\to0$. This result means that the atomic array under the Bragg resonance condition can act as a perfect mirror for the $y$-polarized guided light fields in the limit $N\to\infty$. The loss due to the scattering into the radiation modes is suppressed due to the collective enhancement of scattering into the backward guided modes.
In the limit $NR\ll1$, Eqs.~\eqref{e78} yield $R_N\simeq NR+N(N-1)R^2$ and $T_N\simeq (-1)^{(N+1)n}[T+(N-1)RT+(N-1)^2R^2T]$. The last terms in these expressions contain $N^2$. They are signatures of the collective effects.

\subsubsection{Dependence on the array period}

\begin{figure}[tbh]
\begin{center}
  \includegraphics{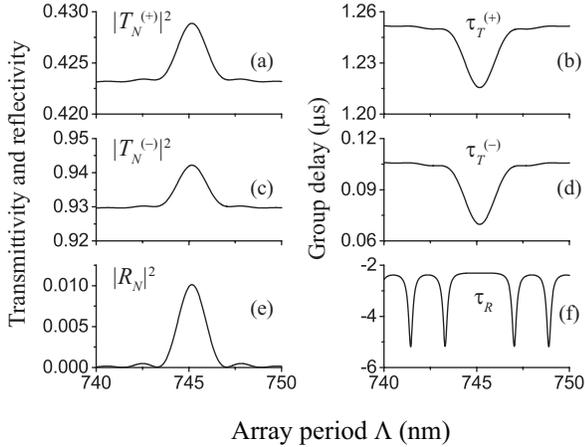}
\end{center}
\caption{
Dependences of the transmittivities $|T_N^{(+)}|^2$ and $|T_N^{(-)}|^2$, the reflectivity $|R_N|^2$, 
and the group delays $\tau_T^{(+)}$, $\tau_T^{(-)}$, and $\tau_R$ on the array period $\Lambda$ in the case of the $x$-polarization scheme. The detuning of the probe field is $\Delta=0$. The number of atoms in the array is $N=200$. Other parameters are as for Fig.~\ref{fig2}.
}
\label{fig20}
\end{figure}

\begin{figure}[tbh]
\begin{center}
  \includegraphics{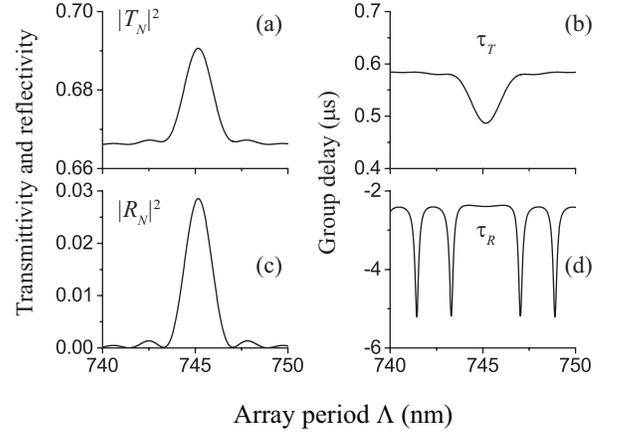}
\end{center}
\caption{
Dependences of the transmittivity $|T_N|^2$, the reflectivity $|R_N|^2$, and the group delays $\tau_T$ and $\tau_R$ on the array period $\Lambda$ in the case of the $y$-polarization scheme. The detuning of the probe field is $\Delta=0$. The number of atoms in the array is $N=200$. Other parameters are as for Fig.~\ref{fig3}.
}
\label{fig21}
\end{figure}

We consider the effect of the array period $\Lambda$ on the Bragg resonance. For this purpose, we plot in Figs.~\ref{fig20} and \ref{fig21} the dependences of the transmittivity $|T_N^{(f)}|^2$, the reflectivity $|R_N|^2$, the transmitted-field group delay $\tau_T^{(f)}$, and the reflected-field group delay $\tau_R$ on the array period $\Lambda$.

The figures show that both the transmittivity and the reflectivity have a local maximum at the array period $\Lambda=745.16$ nm, which satisfies the Bragg resonance condition for the field frequency $\omega_p=\omega_0$. The coexistence of the local maxima of the transmittivity and reflectivity at the Bragg resonance is an interesting feature. This result indicates that the scattering from the atoms into the radiation modes is suppressed due to the Bragg resonance, in agreement with the results of Ref.~\cite{AtomArray}. It is clear that, in the vicinity of the Bragg resonance, the reflectivity for the $y$-polarized guided field [see Fig.~\ref{fig21}(c)] is larger than that for the $x$-polarized guided field [see Fig.~\ref{fig20}(e)]. 
We note that the linewidth of the dependence of the reflectivity and transmittivity on the array period $\Lambda$ is on the order
of a few nanometers. The reason is that, when $\beta_p\delta\Lambda\ll1$, that is, when $\delta\Lambda\ll \lambda_F\equiv2\pi/\beta_p\simeq 745.16$ nm for $\lambda_p=\lambda_0=852.35$ nm, there is no significant deviation from the Bragg resonance condition $\beta_p \Lambda=n\pi$, with $n=1,2,\dots$. Here, $\delta\Lambda$ is a small deviation of the array period $\Lambda$ from a Bragg resonant value.
We observe narrow dips in the $\Lambda$ dependence of $\tau_R$ [see Figs.~\ref{fig20}(f) and \ref{fig21}(d)]. These dips occur at the points where $|R_N|^2$ have minima [see Figs.~\ref{fig20}(e) and \ref{fig21}(c)]. In other words, the dips correspond to the interference fringes that surround the Bragg resonance. The separation $\Delta\Lambda$ between the positions of the dips is determined by the half-period of the function $\sin N\zeta$, which appears in Eqs.~\eqref{e67}. Using the approximation $\zeta\simeq \beta_p\Lambda$, we obtain the estimate $\Delta\Lambda\simeq\pi/N\beta_p$.

\subsubsection{Dependence on the control field intensity}

\begin{figure}[tbh]
\begin{center}
  \includegraphics{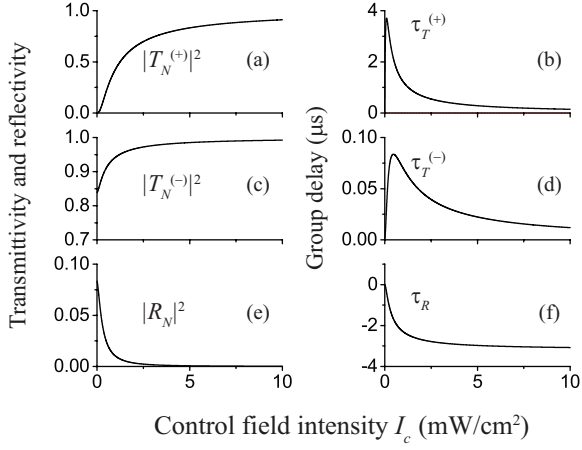}
\end{center}
\caption{
Dependences of the transmittivities $|T_N^{(+)}|^2$ and $|T_N^{(-)}|^2$, the reflectivity $|R_N|^2$, and the group delays $\tau_T^{(+)}$, $\tau_T^{(-)}$, and $\tau_R$ on the control field intensity $I_c$ in the case of the $x$-polarization scheme. The period of the array is $\Lambda=745.16$ nm, which satisfies the second-order Bragg resonance condition. The detuning of the probe field is $\Delta=0$. The number of atoms in the array is $N=200$.
Other parameters are as for Fig.~\ref{fig2}.
}
\label{fig22}
\end{figure}

\begin{figure}[tbh]
\begin{center}
  \includegraphics{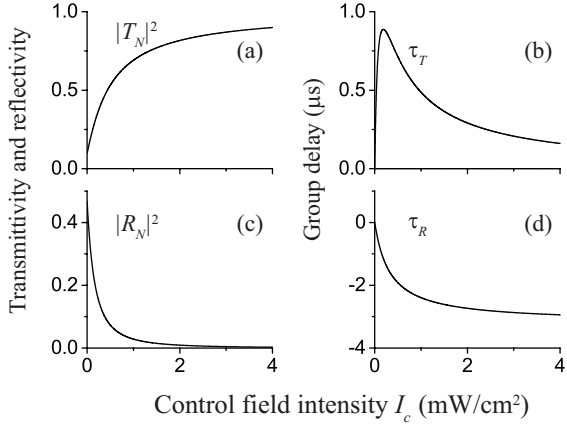}
\end{center}
\caption{
Dependences of the transmittivity $|T_N|^2$, the reflectivity $|R_N|^2$, and the group delays $\tau_T$ and $\tau_R$ on the control field intensity $I_c$ in the case of the $y$-polarization scheme. The period of the array is $\Lambda=745.16$ nm, which satisfies the second-order Bragg resonance condition. The detuning of the probe field is $\Delta=0$. The number of atoms in the array is $N=200$. Other parameters are as for Fig.~\ref{fig3}.
}
\label{fig23}
\end{figure}

In order to show the effect of the control field on the transmission and reflection of the probe field, we plot in Figs.~\ref{fig22} and \ref{fig23} the transmittivity $|T_N^{(f)}|^2$, the reflectivity $|R_N|^2$, the transmitted-field group delay $\tau_T^{(f)}$, and the reflected-field group delay $\tau_R$ as functions of the intensity $I_c$ of the control field $\boldsymbol{\mathcal{E}}_c$. 

The left columns of these figures show that, when $I_c$ increases, the transmittivity $|T_N^{(f)}|^2$ increases but the reflectivity $|R_N|^2$ decreases. 
We observe from the right columns of Figs.~\ref{fig22} and \ref{fig23} that, when $I_c$ increases and is not too small, the group delay $\tau_T^{(f)}$ of the transmitted field decreases, and the absolute value $|\tau_R|$ of the negative group delay $\tau_R<0$ of the reflected field increases. 
The above-mentioned features are the results of the dispersion properties in the EIT window.

\subsubsection{Slow transmitted light and fast reflected light}

\begin{figure}[tbh]
\begin{center}
  \includegraphics{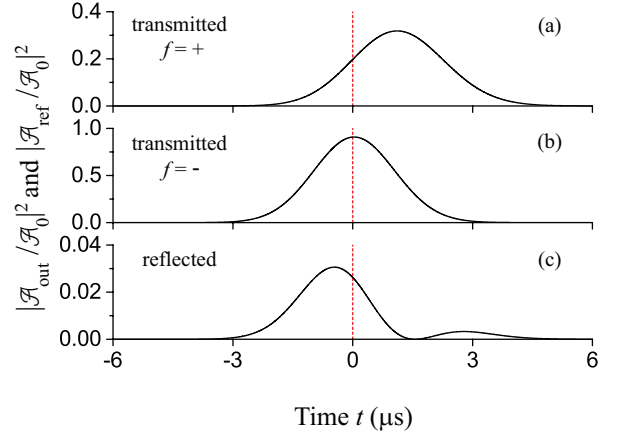}
\end{center}
\caption{(Color online)
Time dependences of the normalized intensities $|\mathcal{A}_{\mathrm{out,ref}}/\mathcal{A}_0|^2$ of the transmitted and reflected pulses in the case of the $x$-polarization scheme. The input pulse is of the Gaussian form, with the central frequency $\omega_p=\omega_0$, the pulse length $\tau_0=2$ $\mu$s, and the peak time $t=0$. The number of atoms is $N=200$. The array period is $\Lambda=745.16$ nm, which satisfies the second-order Bragg resonance condition. Other parameters used are as for Fig.~\ref{fig2}. The dotted red line is for the input-pulse peak time $t=0$ and is a guide to the eye.
}
\label{fig24}
\end{figure}

\begin{figure}[tbh]
\begin{center}
  \includegraphics{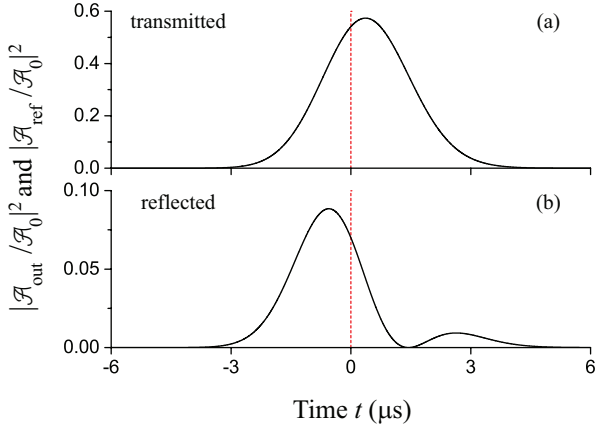}
\end{center}
\caption{(Color online)
Time dependences of the normalized intensities $|\mathcal{A}_{\mathrm{out,ref}}/\mathcal{A}_0|^2$ of the transmitted and reflected pulses in the case of the $y$-polarization scheme. The input pulse is of the Gaussian form, with the central frequency $\omega_p=\omega_0$, the pulse length $\tau_0=2$ $\mu$s, and the peak time $t=0$. The number of atoms is $N=200$. The array period is $\Lambda=745.16$ nm, which satisfies the second-order Bragg resonance condition. Other parameters used are as for Fig.~\ref{fig2}. The dotted red line is for the input-pulse peak time $t=0$ and is a guide to the eye.
}
\label{fig25}
\end{figure}

We plot in Figs.~\ref{fig24} and \ref{fig25} the time dependences of the normalized intensities 
$|\mathcal{A}_{\mathrm{out,ref}}/\mathcal{A}_0|^2$ of the transmitted/reflected pulses in the cases of the $x$- and 
$y$-polarization schemes, respectively. The parameters used are as for Figs.~\ref{fig12} and \ref{fig13} except for the array period $\Lambda=745.16$ nm, which satisfies the second-order Bragg resonance condition.

We observe that, when the probe pulse is $x$ polarized and incident onto the array in the direction $f=+$ [see Fig.~\ref{fig24}(a)] and when the probe pulse is $y$ polarized an incident in an arbitrary propagation direction $f=\pm$ [see Fig.~\ref{fig25}(a)], the transmitted pulse is significantly weakened and delayed by the atomic array. For $N=200$ atoms in the array with the length $L=(N-1)\Lambda\simeq 148$ $\mu$m, we obtain the group delays of about $1.11$ $\mu$s in the case of Fig.~\ref{fig24}(a) and $0.37$ $\mu$s in the case of Fig.~\ref{fig25}(a). These group delays correspond to the group-velocity reduction factors $c/V_g\simeq 2.2\times 10^6$ and $\simeq 0.75\times 10^6$ in the cases of Figs.~\ref{fig24}(a) and \ref{fig25}(a), respectively. Figure \ref{fig24}(b) shows that, when the probe pulse is $x$ polarized and incident onto the array in the direction $f=-$, the intensity reduction and the group delay of the transmitted pulse are not significant. According to Figs. \ref{fig24}(c) and \ref{fig25}(b), the reflected pulse has a negative group delay (fast light) and its intensity is substantial due to the Bragg resonance condition.

Comparison between Figs.~\ref{fig24} and \ref{fig12} and between Figs.~\ref{fig25} and \ref{fig13} shows that the shapes and group delays of the transmitted and reflected pulses are not affected much by the Bragg resonance condition.
However, the intensity of the reflected pulse in the presence of a Bragg resonance [see Figs. \ref{fig24}(c) and \ref{fig25}(b)] is much higher than that in the absence of a Bragg resonance [see Figs. \ref{fig12}(c) and \ref{fig13}(b)].

\subsection{Photonic band gaps}
\label{subsec:gap}

\begin{figure}[tbh]
\begin{center}
  \includegraphics{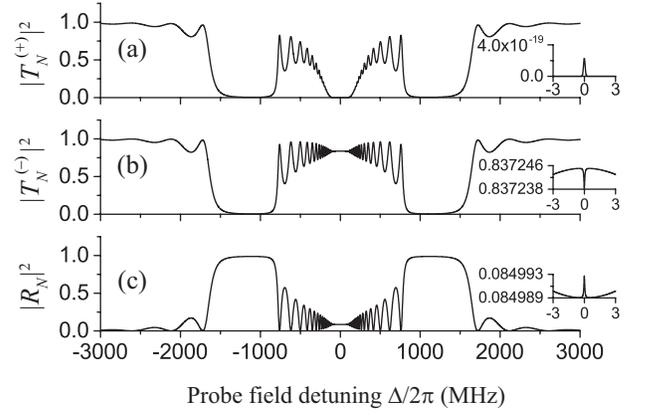}
 \end{center}
\caption{
Photonic band gaps in the case of the $x$-polarization scheme with $\Delta\Lambda=0$.
The transmittivities $|T_N^{(+)}|^2$ (a) and $|T_N^{(-)}|^2$ (b) and the reflectivity $|R_N|^2$ (c) are plotted as functions of the detuning $\Delta$.
The period of the array is $\Lambda=745.16$ nm, which satisfies the second-order Bragg resonance condition. 
The number of atoms in the array is $N=200\,000$. 
The intensity of the control field is $I_c=20$ mW/cm$^2$ (the corresponding Rabi frequency is $\Omega_c=2.06\gamma_0$). 
Other parameters are as for Fig.~\ref{fig2}.
The insets show the narrow structures around the point $\Delta=0$.
}
\label{fig26}
\end{figure}

\begin{figure}[tbh]
\begin{center}
  \includegraphics{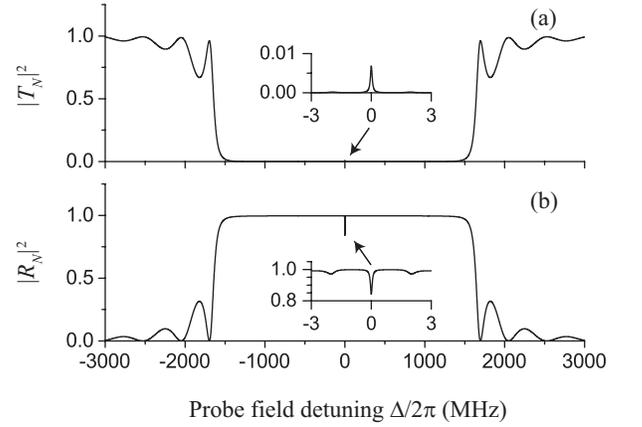}
 \end{center}
\caption{
Photonic band gaps in the case of the $y$-polarization scheme with $\Delta\Lambda=0$.
The transmittivity $|T_N|^2$ (a) and the reflectivity $|R_N|^2$ (b) are plotted as functions of the detuning $\Delta$.
The period of the array is $\Lambda=745.16$ nm, which satisfies the second-order Bragg resonance condition. 
The number of atoms in the array is $N=200\,000$. 
The intensity of the control field is $I_c=20$ mW/cm$^2$ (the corresponding Rabi frequency is $\Omega_c=1.94\gamma_0$). 
Other parameters are as for Fig.~\ref{fig3}.
The insets show the narrow structures around the point $\Delta=0$.
}
\label{fig27}
\end{figure}

It is known that, in the neighborhood of a Bragg resonance, where $\Lambda=\Lambda_{\mathrm{res}}=n\lambda_F/2$ with $n=1,2,\dots$, 
band gaps may be formed when the number of atoms $N$ in the array is large enough \cite{Deutsch95,Chang12,AtomArray}. In order to show the band gaps, we plot in Figs.~\ref{fig26} and \ref{fig27} the transmittivity $|T_N^{(f)}|^2$ and the reflectivity $|R_N|^2$ of the atomic array with the period $\Lambda=745.16$ nm (corresponding to
$\Delta\Lambda\equiv \Lambda-\Lambda_{\mathrm{res}}= 0$ for $\lambda_p=\lambda_0$) as functions of the field detuning $\Delta$ for a very large number of atoms, namely for $N=200\,000$. In the calculations for these figures, we used a large value for the control field intensity $I_c$, namely $I_c=20$ mW/cm$^2$, in order to show clearly the tiny features of the transmittivity and reflectivity for the probe field in the EIT window. Other parameters are as for Figs.~\ref{fig2} and \ref{fig3}. 

Figure~\ref{fig26} shows the frequency dependences of $|T_N^{(+)}|^2$, $|T_N^{(-)}|^2$, and $|R_N|^2$  for the $x$-polarization scheme in the vicinity of a Bragg resonance $\Delta\Lambda=0$. We observe from this figure that, in addition to a narrow plateau around the atomic resonance $\Delta=0$, there are two wide plateaus, one on the left side and the other one on the right side. The left- and right-side plateaus are the photonic band gaps that extend over the frequency range from $\omega_c-\Delta_{\mathrm{max}}$ to $\omega_c-\Delta_{\mathrm{min}}$ and from $\omega_c+\Delta_{\mathrm{min}}$ to $\omega_c+\Delta_{\mathrm{max}}$ \cite{Deutsch95,AtomArray}. 
Here, we have introduced the notations $\omega_c=(\omega_0+\omega_{\mathrm{lat}})/2$ and 
\begin{eqnarray}\label{x78}
\Delta_{\mathrm{max}}&=&\sqrt{\frac{\delta_{\mathrm{lat}}^2}{4}+\frac{u_0|e_r|^2v_g}{\Lambda}},
\nonumber\\
\Delta_{\mathrm{min}}&=&\sqrt{\frac{\delta_{\mathrm{lat}}^2}{4}+\frac{u_0|e_z|^2v_g}{\Lambda}},
\end{eqnarray}
with $\delta_{\mathrm{lat}}=\omega_{\mathrm{lat}}-\omega_0$, and $u_0=\omega_pd_{eg}^2/\epsilon_0\hbar v_g$. The Bragg resonant frequency $\omega_{\mathrm{lat}}$ is determined by the equation $\beta(\omega_{\mathrm{lat}})\Lambda=n\pi$, where the integer number $n=1,2,\dots$ is the order of the Bragg resonance. In deriving Eqs.~\eqref{x78} for the edges of the band gaps, 
we have neglected the atomic decay rate $\Gamma_{eg}$ and the control field Rabi frequency $\Omega_c$. 
In the case of $|\delta_{\mathrm{lat}}|\ll \sqrt{u_0|e_{r,z}|^2v_g/\Lambda}$,
we find that the band gaps will be formed when $N\gg N_{\mathrm{gap}}$, where $N_{\mathrm{gap}}=\sqrt{v_g/u_0\Lambda}/(|e_r|-|e_z|)$. We obtain from Eqs.~\eqref{x78} the estimates $\Delta_{\mathrm{min}}=2\pi\times 847$ MHz and $\Delta_{\mathrm{max}}=2\pi\times 1544$ MHz for the parameters of Fig.~\ref{fig26}. In the band gap regions of Fig.~\ref{fig26}, we have $|R_N|^2\simeq1$ and $|T_N^{(\pm)}|^2\simeq0$. In the central plateau, we have $|R_N|^2\simeq0.085<1$, $|T_N^{(+)}|^2\simeq0$, and $|T_N^{(-)}|^2\simeq0.84$, in agreement with Eqs.~\eqref{x60} and \eqref{e76}. The limiting values $|T_\infty^{(+)}|^2\simeq0$ and $|T_\infty ^{(-)}|^2\simeq0.84\not=0$ of the transmittivities in the central plateau area indicate that the atomic array can operate as an optical diode even in the limit of an infinitely large value of $N$.
The insets of Figs.~\ref{fig26}(a)--\ref{fig26}(c) show that there are narrow structures in the frequency dependences of the transmittivity $|T_N^{(f)}|^2$ and the reflectivity $|R_N|^2$ in
the vicinity of the atomic resonance $\Delta=0$. These tiny structures are the consequences of the interplay between the scattering into the guided and radiation modes, the EIT effect, and the Bragg resonance. 

Figure~\ref{fig27} shows the frequency dependences of $|T_N|^2$ and $|R_N|^2$ for the $y$-polarization scheme in the vicinity of a Bragg resonance $\Delta\Lambda=0$. We observe that there is a wide plateau around the atomic resonance $\Delta=0$. This plateau corresponds to the set of two photonic band gaps that extend over the frequency range from $\omega_c-\Delta_{\mathrm{max}}$ to $\omega_c+\Delta_{\mathrm{max}}$, with frequencies between the atomic frequency $\omega_0$ and 
the frequency $\omega_{\mathrm{lat}}$ excluded \cite{Deutsch95,AtomArray}. Here, we have introduced the notation
\begin{equation}\label{x85}
\Delta_{\mathrm{max}}=\sqrt{\frac{\delta_{\mathrm{lat}}^2}{4}+\frac{2u_0|e_\varphi|^2v_g}{\Lambda}}.
\end{equation}
In deriving Eqs.~\eqref{x85} for the edges of the band gaps, we have neglected the atomic decay rate $\Gamma_{eg}$ and the control field Rabi frequency $\Omega_c$. 
In the case of $|\delta_{\mathrm{lat}}|\ll \sqrt{u_0|e_{\varphi}|^2v_g/\Lambda}$, we find that the band gaps will be formed when $N\gg N_{\mathrm{gap}}$, where $N_{\mathrm{gap}}=\sqrt{2v_g/3u_0\Lambda}/|e_{\varphi}|$. We obtain from Eq.~\eqref{x85} the estimate $\Delta_{\mathrm{max}}=2\pi\times 1561$ MHz for the parameters of Fig.~\ref{fig27}. In the band gap region  of Figs.~\ref{fig27}, we have $|R_N|^2\simeq1$ and $|T_N|^2\simeq0$, in agreement with Eqs.~\eqref{x66a}. The narrow structures in the insets of Figs.~\ref{fig27}(a) and \ref{fig27}(b) are the consequences of the interplay between the scattering into the guided and radiation modes, the EIT effect, and the Bragg resonance.

\begin{figure}[tbh]
\begin{center}
  \includegraphics{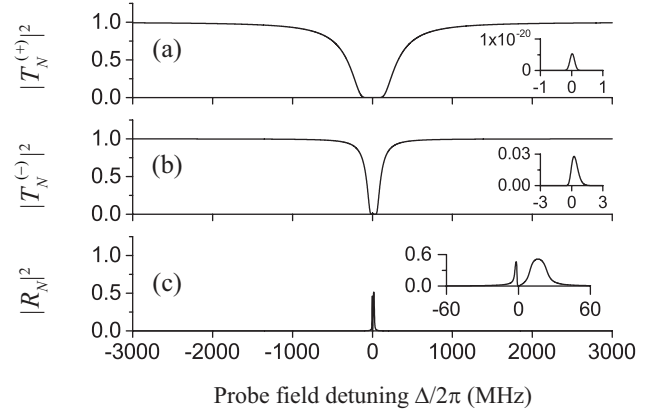}
 \end{center}
\caption{
As Fig.~\ref{fig26} but for $\Delta\Lambda=0.3$ nm. 
}
\label{fig28}
\end{figure}

\begin{figure}[tbh]
\begin{center}
  \includegraphics{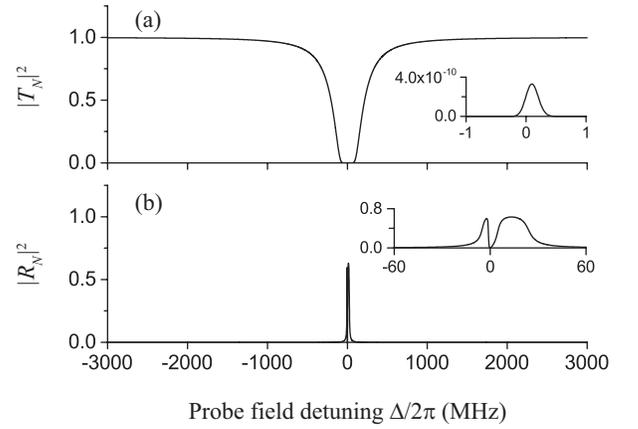}
 \end{center}
\caption{
As Fig.~\ref{fig27} but for $\Delta\Lambda=0.3$ nm. 
}
\label{fig29}
\end{figure}

The band gaps illustrated in Figs.~\ref{fig26} and \ref{fig27} may appear even when $\Omega_c=0$, that is, when there is no control field \cite{Deutsch95,Chang12,AtomArray}.
These band gaps are not related to the EIT effect. We call them the non-EIT band gaps.
When the condition 
\begin{equation} 
|\delta_{\mathrm{lat}}|\gg \sqrt{\frac{u_0|e_{r,\varphi,z}|^2v_g}{\Lambda}} \gg \gamma_0
\end{equation} 
is satisfied, one of the two non-EIT band gaps is far away from the atomic
resonance frequency $\omega_0$ while the other one is close to $\omega_0$. We plot in Figs.~\ref{fig28} and \ref{fig29} the transmittivity $|T_N^{(f)}|^2$ and the reflectivity $|R_N|^2$ of the atomic array as functions of the field detuning $\Delta$ in the case of $\Lambda=745.46$ nm, that is, $\Delta\Lambda\equiv \Lambda-\Lambda_{\mathrm{res}}=0.3$ nm, which corresponds to $\delta_{\mathrm{lat}}=-v_g\beta_p^2\Delta\Lambda/n\pi\simeq-107$ GHz.
Other parameters are the same as for Figs.~\ref{fig26} and \ref{fig27}. We observe from the insets of Figs.~\ref{fig28}(c) and \ref{fig29}(b) 
that, in the vicinity of $\Delta=0$, there are two peak regions where the reflectivity $|R_N|^2$ is significant \cite{Petrosyan07,Schilke12}. 
One of these peak regions is broad and is a non-EIT band gap. The other peak region is narrow and is also a band gap. The occurrence of this additional band gap is due to the EIT effect caused by the action of the control field $\boldsymbol{\mathcal{E}}_c$ \cite{Petrosyan07,Schilke12}. 

In the above calculations, we did not include the experimental limited filling ratio of the atomic array. In the Vienna atom trap experiment \cite{Vetsch10}, due to the small trapping volumes, the loading operated in the collisional blockade
regime \cite{Schlosser02}. This results in an occupancy of at most one atom per trapping site. 
For the parameters of the experiment \cite{Vetsch10}, the filling ratio is about 0.5 \cite{Vetsch10,Schlosser02}.
We note that, in the case where the array period $\Lambda$ is far from the Bragg resonance, the inclusion of the filling ratio will lead to just a reduction of the atom-number density $n_A$ and an increase of the medium length $L$ for a given atom number $N$. In the case where the array period $\Lambda$ is near to a Bragg resonance, the presence of a void in the array will lead to just an additional phase of $n\pi$ for the transmission coefficient $T^{(f)}_N$, where $n$ is the order of the Bragg resonance. Therefore, we expect that the inclusion of the filling ratio will not affect the results for the above limiting cases.
In the intermediate case where the array period is not far from and not close to a Bragg resonance, the inclusion of the filling ratio will lead to a reduction of the reflectivity (and possibly also a reduction of the transmittivity) of the array with a given atom number $N$. In particular, the widths of the resonances in Figs.~\ref{fig20} and \ref{fig21} will be reduced.

\section{Summary}
\label{sec:summary}

We have studied the propagation of guided light along an array of three-level atoms trapped in the vicinity of an optical nanofiber under the EIT condition. We have examined two schemes of atomic levels and field polarizations where the guided probe field is quasilinearly polarized along the major principal axis $x$ or the minor principal axis $y$. We have derived the coupled-mode propagation equations, the input-output relation, the scattering matrix, and the transfer matrix for the transmitted and reflected fields. We have taken into account the complexity of the polarization of the guided field and the discreteness and periodicity of the atomic positions in the array. We have calculated the reflection and transmission coefficients. We have found that, when the array period is far from the Bragg resonance, the reflection is negligible and the homogeneous-medium approximation can be used. We have numerically demonstrated that 200 cesium atoms in a linear array with a length of 100 $\mu$m at a distance of 200 nm from the surface of a nanofiber with a radius of 250 nm can slow down the speed of guided probe light by a factor of about $3.5\times 10^6$ (the group delay is about 1.17 $\mu$s). In the neighborhood of the Bragg resonance, a significant fraction of the guided probe light can be reflected back with a negative group delay (that is, with a positive group advance). The reflectivity and the group delay of the reflected field do not depend on the propagation direction of the probe field. However, when the input guided light is quasilinearly polarized along the major principal axis $x$, the transmittivity and the group delay of the transmitted field substantially depend on the propagation direction of the probe field. When the input guided light is quasilinearly polarized along the major principal axis $x$ and propagates in the direction $f=-\mathrm{sign}(M_e-M_g)$, under the Bragg resonance condition for an array of atoms in an appropriate internal state, the transmission of the guided light is not zero even in the limit of large atom number $N$. This result indicates that the atomic array can operate as an optical diode even in the limit of infinitely large atom numbers \cite{Sayrin15b}. The directionality of transmission of guided light through the atomic array is a consequence of the existence of a longitudinal component of the guided light field as well as the ellipticity of both the field polarization and the atomic dipole vector.

\begin{acknowledgments}
We thank P. Schneeweiss for helpful comments and discussions.
F.L.K. acknowledges support by the Austrian Science Fund (Lise Meitner Project No. M 1501-N27)
and by the European Commission (Marie Curie IIF Grant No. 332255). 
\end{acknowledgments}



\begin{thebibliography}{99}

\bibitem{Harris review} S. E. Harris, Phys. Today \textbf{50}, No. 7, 36 (1997). 

\bibitem{Scully review} M. O. Scully, Phys. Rep. \textbf{219}, 191 (1992).

\bibitem{slow light review} For reviews see M. D. Lukin, P. Hemmer, and M. O. Scully, Adv. At.,
Mol., Opt. Phys. \textbf{42}, 347 (2000); N. V. Vitanov, M. Fleischhauer, B. W. Shore, and K. Bergmann, \textit{ibid.} \textbf{46}, 55 (2001);
A. B. Matsko, O. Kocharovskaya, Y. Rostovtsev, G. R. Welch, A. S. Zibrov, and M. O. Scully, \textit{ibid.} \textbf{46}, 191 (2001); 
R. W. Boyd and D. J. Gauthier, Prog. Opt. \textbf{43}, 497 (2002); M. D. Lukin, Rev. Mod. Phys. \textbf{75}, 457 (2003). 

\bibitem{most recent review} For a more recent review see
M. Fleischhauer, A. Imamoglu, and J. P. Marangos, Rev. Mod. Phys. \textbf{77}, 633 (2005).

\bibitem{Scully} M. O. Scully and M. S.  Zubairy, \textit{Quantum Optics} (Cambridge University Press, New York, 1997).

\bibitem{Agarwal book} G. S. Agarwal, \textit{Quantum Optics} (Cambridge University Press, New York, 2013).

\bibitem{vapor} A. Kasapi, M. Jain, G. Y. Yin, and S. E. Harris, Phys. Rev. Lett. \textbf{74}, 2447 (1995); M. M. Kash, V. A. Sautenkov, A. S. Zibrov, L. Hollberg, G. R. Welch, M. D. Lukin, Y. Rostovtsev, E. S. Fry, and M. O. Scully, \textit{ibid.} \textbf{82}, 5229 (1999);
D. Budker, D. F. Kimball, S. M. Rochester, and V. V. Yashchuk, \textit{ibid.} \textbf{83}, 1767 (1999);
D. F. Phillips, A. Fleischhauer, A. Mair, R. L. Walsworth, and M. D. Lukin, \textit{ibid.} \textbf{86}, 783 (2001); A. S. Zibrov, A. B. Matsko, O. Kocharovskaya, Y. V. Rostovtsev, G. R. Welch, and M. O. Scully, \textit{ibid.} \textbf{88}, 103601 (2002).

\bibitem{ultracold} L. V. Hau, S. E. Harris, Z. Dutton, and C. Behroozi, Nature (London) \textbf{397}, 594 (1999); C. Liu, Z. Dutton, C. Behroozi, and L. V. Hau, \textit{ibid.} \textbf{409}, 490 (2001).

\bibitem{polariton} M. Fleischhauer and M. D. Lukin, Phys. Rev. Lett. \textbf{84}, 5094 (2000).

\bibitem{Shen} J. Q. Shen and S. He, Phys. Rev. A \textbf{74}, 063831 (2006); J. Q. Shen, \textit{ibid.} \textbf{84}, 063841 (2011); J. Q. Shen, J. Phys. Soc. Jpn. \textbf{81}, 024403 (2012).  

\bibitem{Andre} A. Andr\'{e}, M. Bajcsy, A. S. Zibrov, and M. D. Lukin, Phys. Rev. Lett. \textbf{94}, 063902 (2005). 

\bibitem{Ghosh} S. Ghosh, A. R. Bhagwat, C. K. Renshaw, S. Goh, A. L. Gaeta, and B. J. Kirby,
Phys. Rev. Lett. \textbf{97}, 023603 (2006).

\bibitem{Bajcsy09} M. Bajcsy, S. Hofferberth, V. Balic, T. Peyronel, M. Hafezi, A. S. Zibrov, V. Vuletic, and M. D. Lukin, Phys. Rev. Lett. \textbf{102}, 203902 (2009).

\bibitem{Neff} C. W. Neff, L. M. Andersson, and M. Qiu, Opt. Express \textbf{15}, 10362 (2007). 
 
\bibitem{Li} T. Li, H. Wang, N. H. Kwong, and R. Binder, Opt. Express \textbf{11}, 3298 (2003); H. Wang and S. O'Leary, J. Opt. Soc. Am. B \textbf{29}, A6 (2012).

\bibitem{Bermel} P. Bermel, A. Rodriguez, S. G. Johnson, J. D. Joannopoulos, and M. Soljacic, Phys. Rev. A \textbf{74}, 043818 (2006); X. Yu, J. Li, and X. Li, J. Opt. Soc. Am. B \textbf{30}, 649 (2013). 

\bibitem{Patnaik} A. K. Patnaik, J. Q. Liang, and K. Hakuta, Phys. Rev. A \textbf{66}, 063808 (2002).

\bibitem{propag} Fam Le Kien and K. Hakuta, Phys. Rev. A \textbf{79}, 013818 (2009).

\bibitem{fibercavity} Fam Le Kien and K. Hakuta, Phys. Rev. A \textbf{79}, 043813 (2009).

\bibitem{Spillane} S. M. Spillane, G. S. Pati, K. Salit, M. Hall, P. Kumar, R. G. Beausoleil, 
and M. S. Shahriar, Phys. Rev. Lett. \textbf{100}, 233602 (2008). 

\bibitem{Laurat15} B. Gouraud, D. Maxein, A. Nicolas, O. Morin, and J. Laurat, arXiv:1502.01458. 

\bibitem{Sayrin15a} C. Sayrin, C. Clausen, B. Albrecht, P. Schneeweiss, and A. Rauschenbeutel, Optica \textbf{2}, 353 (2015). 

\bibitem{Ginzburg} P. Ginzburg and M. Orenstein, Opt. Express \textbf{14}, 11312 (2006); C. Ciret, M. Alonzo, V. Coda, A. A. Rangelov, and G. Montemezzani, Phys. Rev. A \textbf{88}, 013840 (2013). 

\bibitem{Hong} T. Hong, Phys. Rev. Lett. \textbf{90}, 183901 (2003); Y. Qi, G. Lin, J. Yang, Y. Niu, and S. Gong, J. Opt. Soc. Am. B \textbf{31}, 445 (2014); C. Hang and G. Huang,
Phys. Rev. A \textbf{89}, 013821 (2014).  

\bibitem{Tarhan} D. Tarhan, N. Postacioglu, and O. E. Mustecaplioglu, Opt. Lett. \textbf{32}, 1038 (2007). 

\bibitem{Mazur's Nature}  L. Tong, R. R. Gattass, J. B. Ashcom, S. He, J. Lou, M. Shen, I. Maxwell, and  E. Mazur, Nature (London) \textbf{426}, 816 (2003).

\bibitem{Birks} T. A. Birks, W. J. Wadsworth, and P. St. J. Russell,  Opt. Lett. \textbf{25}, 1415 (2000); S. G. Leon-Saval, T. A. Birks, W. J. Wadsworth, P. St. J. Russell, and  M. W. Mason,
in \textit{Conference on Lasers and Electro-Optics (CLEO),
Technical Digest}, postconference ed. (Optical Society of America, Washington, D.C., 2004),
paper CPDA6. 

\bibitem{taper} J. C. Knight, G. Cheung, F. Jacques, and T. A. Birks, Opt. Lett. \textbf{22}, 1129 (1997); M. Cai and K. Vahala, \textit{ibid.} \textbf{26}, 884 (2001).

\bibitem{Dawkins11} S. T. Dawkins, R. Mitsch, D. Reitz, E. Vetsch, and A. Rauschenbeutel, 
Phys. Rev. Lett. \textbf{107}, 243601 (2011). 

\bibitem{Reitz13} D. Reitz, C. Sayrin, R. Mitsch, P. Schneeweiss, and A. Rauschenbeutel, Phys. Rev. Lett. \textbf{110}, 243603 (2013).

\bibitem{Reitz14} D. Reitz, C. Sayrin, B. Albrecht, I. Mazets, R. Mitsch, P. Schneeweiss, and A. Rauschenbeutel, Phys. Rev. A \textbf{89}, 031804(R) (2014). 

\bibitem{Mitsch14a} R. Mitsch, C. Sayrin, B. Albrecht, P. Schneeweiss, and A. Rauschenbeutel, Phys. Rev. A \textbf{89}, 063829 (2014).

\bibitem{Mitsch14b} R. Mitsch, C. Sayrin, B. Albrecht, P. Schneeweiss, and A. Rauschenbeutel, Nat. Commun. \textbf{5}, 5713 (2014). 

\bibitem{Sayrin15b} C. Sayrin, C. Junge, R. Mitsch, B. Albrecht, D. O'Shea, P. Schneeweiss, J. Volz, and A. Rauschenbeutel, arXiv:1502.01549.

\bibitem{Vetsch10} E. Vetsch, D. Reitz, G. Sagu\'{e}, R. Schmidt, S. T. Dawkins, and A. Rauschenbeutel, Phys. Rev. Lett. \textbf{104}, 203603 (2010).

\bibitem{Goban12} A. Goban, K. S. Choi, D. J. Alton, D. Ding, C. Lacro\^{u}te, M. Pototschnig, T. Thiele, N. P. Stern, and H. J. Kimble, Phys. Rev. Lett. \textbf{109}, 033603 (2012).

\bibitem{Polzik14}  J. B. B\'{e}guin, E. M. Bookjans, S. L. Christensen, H. L. S{\o}rensen, J. H. M\"{u}ller, E. S. Polzik, and J. Appel,    Phys. Rev. Lett. \textbf{113}, 263603 (2014).

\bibitem{Petersen14} J. Petersen, J. Volz, and A. Rauschenbeutel,  Science \textbf{346}, 67 (2014).

\bibitem{Deutsch95} I. H. Deutsch, R. J. C. Spreeuw, S. L. Rolston, and W. D. Phillips, Phys. Rev. A \textbf{52}, 1394 (1995).

\bibitem{Birkl95} G. Birkl, M. Gatzke, I. H. Deutsch, S. L. Rolston, and W. D. Phillips, Phys. Rev. Lett. \textbf{75}, 2823 (1995). 

\bibitem{Henkel03} G. Boedecker and C. Henkel, Opt. Express \textbf{11}, 1590 (2003).

\bibitem{Artoni05} M. Artoni, G. La Rocca, and F. Bassani, Phys. Rev. E \textbf{72}, 046604 (2005). 

\bibitem{Schilke11} A. Schilke, C. Zimmermann, P. W. Courteille, and W. Guerin, Phys. Rev. Lett. \textbf{106}, 223903 (2011).

\bibitem{Chang07} D. E. Chang, A. S. S{\o}rensen, E. A. Demler, and M. D. Lukin, Nature Phys. \textbf{3}, 807 (2007).

\bibitem{Chang12} D. E. Chang, L. Jiang, A. V. Gorshkov, and H. J. Kimble, New J. Phys. \textbf{14}, 063003 (2012).

\bibitem{Chang11} Y. Chang, Z. R. Gong, and C. P. Sun, Phys. Rev. A \textbf{83}, 013825 (2011).

\bibitem{Petrosyan07} D. Petrosyan, Phys. Rev. A \textbf{76}, 053823 (2007).

\bibitem{Schilke12} A. Schilke, C. Zimmermann, and W. Guerin, Phys. Rev. A \textbf{86}, 023809 (2012).

\bibitem{Ritsch14a} S. Ostermann, M. Sonnleitner, and H. Ritsch, New J. Phys. \textbf{16}, 043017 (2014).

\bibitem{Ritsch14b} D. Holzmann, M. Sonnleitner, and H. Ritsch, Eur. Phys. J. D \textbf{68}, 352 (2014).

\bibitem{AtomArray} Fam Le Kien and A. Rauschenbeutel,  Phys. Rev. A \textbf{90}, 063816 (2014). 

\bibitem{Tsoi09} T. S. Tsoi and C. K. Law, Phys. Rev. A \textbf{80}, 033823 (2009).

\bibitem{fiber books} See, for example, D. Marcuse, \textit{Light Transmission Optics} (Krieger, Malabar, FL,  1989); K. Okamoto, \textit{Fundamentals of Optical Waveguides} (Academic Press, New York, 2006); A. W. Snyder and J. D. Love, \textit{Optical Waveguide Theory} (Chapman and Hall, New York, 1983).

\bibitem{fibermode} Fam Le Kien, J. Q. Liang, K. Hakuta, and V. I. Balykin, 
Opt. Commun. \textbf{242}, 445 (2004); L. Tong, J. Lou, and  E. Mazur, Opt. Express \textbf{12}, 1025 (2004).

\bibitem{Fam14} Fam Le Kien and A. Rauschenbeutel, Phys. Rev. A \textbf{90}, 023805 (2014). 

\bibitem{cesium decay} Fam Le Kien, S. Dutta Gupta, V. I. Balykin, and K. Hakuta, 
Phys. Rev. A \textbf{72}, 032509 (2005).

\bibitem{Steck02} D.~A.~Steck, {\em Cesium D Line Data} (23 January 1998, Revision 2.1.4, 23 December 2010), http://steck.us/alkalidata/.

\bibitem{coolingbook} H. J. Metcalf and P. van der Straten, \textit{Laser Cooling and Trapping} (Springer, New York, 1999).

\bibitem{Boyd09} R. W. Boyd and D. J. Gauthier, Science \textbf{326}, 1074 (2009).

\bibitem{Wang00} L. J. Wang, A. Kuzmich, and  A. Dogariu,  Nature (London) \textbf{406}, 277 (2000).
 
\bibitem{Schlosser02} N. Schlosser, G. Reymond, and P. Grangier, Phys. Rev. Lett. \textbf{89}, 023005 (2002).

\end{thebibliography}
\end{document}